\documentclass[usenatbib]{mnras}
\usepackage{graphicx}
\usepackage{pdflscape}
\usepackage{newtxtext,newtxmath}
\usepackage[figuresright]{rotating} 
\usepackage{caption}
\usepackage{multirow} 
\usepackage{booktabs} 
\usepackage{ulem} 
\usepackage[T1]{fontenc}
\usepackage{ae,aecompl}
\usepackage[dvipsnames]{xcolor}

\title[Activity and magnetic field of late-type stars]{Linking chromospheric activity and magnetic field properties for late-type dwarf stars}

\author[E. L. Brown et al.]{E. L. Brown$^{1}$\thanks{E-mail: emma.brown@usq.edu.au},
S. V. Jeffers$^{2}$,
S. C. Marsden$^{1}$,
J. Morin$^{3}$,
S. Boro Saikia$^{4}$,
P. Petit$^{5}$,\newauthor
M. M. Jardine$^{6}$,
V. See$^{7,8}$,
A. A. Vidotto$^{9,10}$,
M. W. Mengel$^{1}$,
M. N. Dahlkemper$^{11,12}$ and\newauthor
the BCool Collaboration $^{13}$
\\
$^{1}$University of Southern Queensland, Centre for Astrophysics, Toowoomba, QLD, 4350, Australia\\
$^{2}$Max Planck Institut for Solar System Research, Justus-von-Liebig-Weg 3, 37077 G\"ottingen, Germany\\
$^3$LUPM, Universit\'e de Montpellier, CNRS, Place Eug\'ene Bataillon, F-34095 Montpellier, France\\
$^{4}$University of Vienna, Department of Astrophysics, Turkenschanzstrasse 17, 1180 Vienna, Austria\\
$^5$Institut de Recherche en Astrophysique et Plan\'etologie, Universit\'e de Toulouse, CNRS, CNES, 31400 Toulouse, France\\
$^6$SUPA, School of Physics and Astronomy, University of St Andrews, North Haugh, St Andrews, KY16 9SS, UK\\
$^7$European Space Agency (ESA), European Space Research and Technology Centre (ESTEC), Keplerlaan 1, 2201 AZ Noordwijk, The Netherlands\\
$^8$University of Exeter, Department of Physics \& Astronomy, Stoker Road, Devon, Exeter, EX4 4QL, UK\\
$^9$School of Physics, Trinity College Dublin, The University of Dublin, College Green, Dublin-2, Ireland\\
$^{10}$Leiden Observatory, Leiden University, PO Box 9513, 2300 RA Leiden, The Netherlands\\
$^{11}$University of G\"ottingen, Physics Education Research, Friedrich Hund Platz 1, 37077, G\"ottingen, Germany\\
$^{12}$CERN, Teacher and Student Programmes, Esplanade de Particules 1, CH-1211, Genève 23, Switzerland\\
$^{13}$https://bcool.irap.omp.eu/}

\date{Accepted XXX. Received YYY; in original form ZZZ}

\pubyear{2021}

\begin{document}
\label{firstpage}
\pagerange{\pageref{firstpage}--\pageref{lastpage}}
\maketitle

\begin{abstract}
Spectropolarimetric data allow for simultaneous monitoring of stellar chromospheric $\log{R^{\prime}_{\rm{HK}}}$ activity and the surface-averaged longitudinal magnetic field, $B_l$, giving the opportunity to probe the relationship between large-scale stellar magnetic fields and chromospheric manifestations of magnetism. We present $\log{R^{\prime}_{\rm{HK}}}$ and/or $B_l$ measurements for 954 mid-F to mid-M stars derived from spectropolarimetric observations contained within the PolarBase database.  Our magnetically active sample complements previous stellar activity surveys that focus on inactive planet-search targets. We find a positive correlation between mean $\log{R^{\prime}_{\rm{HK}}}$ and mean $\log|B_l|$, but for G stars the relationship may undergo a change between $\log{R'_{\rm{HK}}}\sim-4.4$ and $-4.8$. The mean $\log{R^{\prime}_{\rm{HK}}}$ shows a similar change with respect to the $\log{R^{\prime}_{\rm{HK}}}$ variability amplitude for intermediately-active G stars.  We also combine our results with archival chromospheric activity data and published observations of large-scale magnetic field geometries derived using Zeeman Doppler Imaging.  The chromospheric activity data indicate a slight under-density of late-F to early-K stars with $-4.75\leq\log{R'_{\rm HK}}\leq-4.5$. This is not as prominent as the original Vaughan--Preston gap, and we do not detect similar under-populated regions in the distributions of the mean $|B_l|$, or the $B_l$ and $\log{R'_{\rm HK}}$ variability amplitudes. Chromospheric activity, activity variability and toroidal field strength decrease on the main sequence as rotation slows. For G stars, the disappearance of dominant toroidal fields occurs at a similar chromospheric activity level as the change in the relationships between chromospheric activity, activity variability and mean field strength.
\end{abstract}


\DeclareRobustCommand{\DO}[3]{#2}
\DeclareRobustCommand{\DE}[3]{#2}

\section{Introduction}

The dynamo-driven magnetic fields of cool stars generate a range of activity phenomena with widespread impacts.  Magnetized stellar winds influence circumstellar environments, including the formation, evolution and potential habitability of surrounding planets \citep[e.g. ][]{Fionnagain2018}. The detectability and parameterization of these planets is also complicated by the temporal evolution of magnetic features such as faculae and starspots on the stellar surface \citep{Queloz2001,Huelamo2008}. Magnetic activity is  intricately linked to the rotational evolution of a star, since the coupling of the magnetic field with magnetic wind causes the loss of stellar angular momentum \citep{Skumanich1972,Gallet2015,Finley2018,see2019_spindown}. For this reason, activity is often used as a proxy to estimate the rotation periods \citep[i.e.][]{Noyes1984} and ages \citep{Lorenzo-Oliveira2018} of stars.  Thus, improving our understanding of the stellar dynamo, its evolution throughout the life of a star, interplay with stellar rotation and other stellar properties, and the induced surface activity are important goals.  

Stellar dynamo theory is largely informed by observations of the Sun. It is well established that the cyclic, solar magnetic dynamo is powered by convection and rotation within the outer convective layers of the Sun \citep{Charbonneau2005}. Differential rotation is a key mechanism that converts an initially poloidal field to a toroidal field, and then the poloidal field is regenerated through cyclonic convection or the Babcock-Leighton mechanism \citep{Charbonneau2005}. Some features of solar magnetic cycles are still not well explained nor reliably predicted using current dynamo theory, such as the amplitudes and lengths of cycles \citep{Petrovay2020}, so monitoring of the magnetism of other cool stars has become an important adjunct to studies of the Sun. Through long-term monitoring of the magnetic fields and surface activity of late-type stars with a range of properties, we can provide context for the Sun's magnetic behaviour and can work toward untangling the intricate relationships between stellar properties and the magnetic dynamo. 

Chromospheric \ion{Ca}{ii} H\&K spectral line emissions provide an important, indirect diagnostic of stellar magnetic activity. Magnetic heating in the chromosphere causes increased emissions in the cores of the \ion{Ca}{ii} H\&K spectral lines, both in the Sun and other cool stars \citep{Eberhard1913}. \ion{Ca}{ii} H\&K activity is usually represented by the S-index \citep{Vaughan1978}, where the core emissions in the H and K lines are normalised to the continuum. Alternatively, when comparing the activity of stars across spectral types, photospheric and basal flux contributions are removed to provide a purely chromospheric activity index, $\log R^{\prime}_{\rm HK}$.  Decades-long monitoring programs, such as the Mount Wilson Program
\citep{Wilson1978,Duncan1991,Baliunas1995}, have revealed S-index variability on a range of timescales and amplitudes across the HR diagram \citep{Wilson1978,Duncan1991,Baliunas1995,Hall2007,Lehtinen2016,BoroSaikia2018,Gomes2020}, with some stars showing cyclic S-index variability similar to the Sun's 11-year activity cycle. The relationships between mean activity levels, chromospheric activity cycle periods and stellar properties have been previously investigated by  \citet{Noyes1984,Bohm2007,Jenkins2011,Brandenburg2017,BoroSaikia2018} and \citet{Gomes2020}, among others. These studies suggest that cool stars may fall into distinct activity groups based on their mean chromospheric activity and activity cycle periods. \citet{Noyes1984} first reported that main sequence stars fall into `active' and `inactive' groups when their mean $\log R'_{\rm HK}$ is plotted against stellar B-V. They reported an apparent lack of F and G-type stars with intermediate mean chromospheric activity, known as the `Vaughan-Preston gap'. More recently, \citet{BoroSaikia2018} determined that the Vaughan-Preston gap is not prominent in their chromospheric activity study of 4454 cool stars drawn from multiple surveys. \citet{Gomes2020}, in their sample of 1674 FGK stars from the AMBRE-HARPS sample \citep{deLaverny2013}, identified up to three different activity regimes for F-K stars, each separated by reduced populations of stars and occurring at different activity levels for different stellar types. Explanations for the reduced populations between activity groups range from sample biases to phases of rapid stellar evolution. As yet, there is no consensus on the existence of distinct activity groups, nor their physical basis. 

Chromospheric activity variability has been less studied compared to mean activity levels or the periods of activity cycles. 
\citet{Saar2002} studied chromospheric activity cycle amplitudes for 31 stars from the Mount Wilson Project with well-defined chromospheric activity cycles, finding that peak-to-peak cycle amplitudes increase with mean activity. \citet{Gomes2020} also studied the relationship between mean chromospheric activity and activity variability amplitudes for their 1674 AMBRE-HARPS stars, but their data consisted of less-dense time series' of observations, covering shorter time spans, compared to that used by \citet{Saar2002}. \citet{Gomes2020} found that, although the upper envelope of chromospheric activity variability scales with mean activity level, low activity stars do not necessarily have low chromospheric variability amplitudes. This implies that planet-search stars, which are targeted for low chromospheric activity variability, cannot be identified based on their mean chromospheric activity level alone. 
Several studies have also compared the amplitudes of stellar photometric variability to chromospheric variability amplitudes and mean chromospheric activity levels \citep[i.e.][]{Radick1998,Lockwood2007,Radick2018}. Stellar photometric variability amplitudes generally increase with mean chromospheric activity levels, but the Sun appears to have a lower photometric variability compared to other stars with similar mean chromospheric activity \citep{Radick2018,Reinhold2020}. Young, active stars tend to become darker as their chromospheric emissions increase throughout activity cycles, while mature and less active stars, including the Sun, become brighter with increasing chromospheric emissions. This suggests a transition from spot dominated to faculae dominated photometric variability at around middle age \citep{Radick2018}. 

In this paper we analyse time-series spectropolarimetric observations for a sample of main sequence and youthful F-M stars from the PolarBase\footnote{http://polarbase.irap.omp.eu/} database \citep{Petit2014}. Spectropolarimetry uniquely allows for the simultaneous measurement of the strength of the surface-averaged, large-scale stellar magnetic field ($B_l$) and the chromospheric emissions in the cores of the \ion{Ca}{ii} H\&K lines. Thus, the data provide an opportunity to compare large--scale magnetic fields with the properties of smaller--scale magnetic surface features. These are known to be related \citep{Lehmann2018b} but the nature of the relation is not yet clear.  

The PolarBase data span the years 2005 through to 2019, allowing for long-term monitoring of magnetic and chromospheric activity variability. We present both the average chromospheric activity and magnetic field strengths, and their variability amplitudes.  We also combine our data with archival $\log R'_{\rm HK}$ measurements from \citet{Gomes2020} and \citet{BoroSaikia2018}, as well as details of the magnetic field geometries for $\sim 60$ of our PolarBase targets which have been the subjects of previous Zeeman Doppler Imaging \citep[ZDI,][]{Semel1989} studies.

The key aims of this work are to (i) compile a database of chromospheric activity measurements and surface-averaged large-scale magnetic field strengths for cool stars in the PolarBase sample, (ii) compare the distributions of chromospheric activity, $B_l$ and their variability across the sample, (iii) search for distinct activity groups in the activity distribution, as well as any $B_l$ counterparts, and (iv) use published ZDI results to gain new insights into the relationships between chromospheric activity, activity variability, magnetic field geometry and stellar properties.
\section{The stellar sample}

\subsection{Sample selection}

PolarBase contains spectra for >2000 stellar objects, sourced from the high-resolution spectropolarimeters ESPaDOnS \citep{Espadons1,Espadons2}, coupled with the 3.6\,m Canada-France-Hawaii Telescope at Mauna Kea Observatory, and NARVAL \citep{Auri2003}, coupled with the 2\,m Telescope Bernard Lyot at Pic du Midi Observatory. Targets cover the spectral types O4 to M9, with 60 percent of targets being cooler than F5 \citep{Petit2014}. We filtered the PolarBase targets to include only stars that are within the GAIA catalogue \citep[DR2,][]{GAIA2016,GAIA2018}. We included  only stars with effective temperatures between 3200 and 6700 K, which correspond roughly to spectral types M5 through to F5, and aimed to include pre--main sequence and main sequence stars only. We initially filtered giants and more evolved stars from the sample based on their luminosity classification within {\sc{simbad}}\footnote{http://simbad.u-strasbg.fr/simbad/}, where available. We then further filtered out subgiant stars using the method described in section \ref{sec:Li}.  From our filtered target list, we include in this paper only the 954 stars for which we were able to measure the chromospheric $\log{R'_{\rm HK}}$ (section \ref{sec:Rhk}) and/or the longitudinal magnetic field, $B_l$ (sections \ref{sec:LSD} and \ref{sec:B_l}).  A HR diagram of the sample is shown in Figure \ref{fig:HRD}. PolarBase stars are generally targeted because they are magnetically active, so our sample is comparatively more active and complements previous chromospheric activity surveys that derive from samples of planet-search targets \citep[e.g. ][]{Gomes2020}. 

\subsection{Stellar properties}

Stellar properties are shown in Table \ref{tab:stellar_properties}. The full Table \ref{tab:stellar_properties} is available online. 

\subsubsection{Effective temperature, luminosity, surface gravity, radius and B-V}\label{sec:params_from_MIST}

We estimated the effective temperature, luminosity, surface gravity ($\log g$), stellar radius and Tycho $B_T$ and $V_T$ magnitudes by interpolating a grid of MIST (MESA Isochrones \& Stellar Tracks) evolutionary tracks \citep{Mist3,Mist4,Mist5,Mist6,Mist2,Mist1} overlaid on the GAIA colour-magnitude diagram (G versus BP-RP magnitudes) for our sample. BP and RP magnitudes were taken directly from GAIA DR2 \citep{GAIA2016,GAIA2018}, and the absolute G-band magnitudes were derived using the apparent G magnitudes and parallaxes from GAIA DR2.   We assumed solar metallicity \citep[as in e.g.][]{Arun2019}. We converted the Tycho $B_T$ and $V_T$ magnitudes to the Johnson B-V magnitude shown in Table \ref{tab:stellar_properties} according to \citet{tycho}. 

\begin{figure}
    \centering
    \includegraphics[width=\columnwidth]{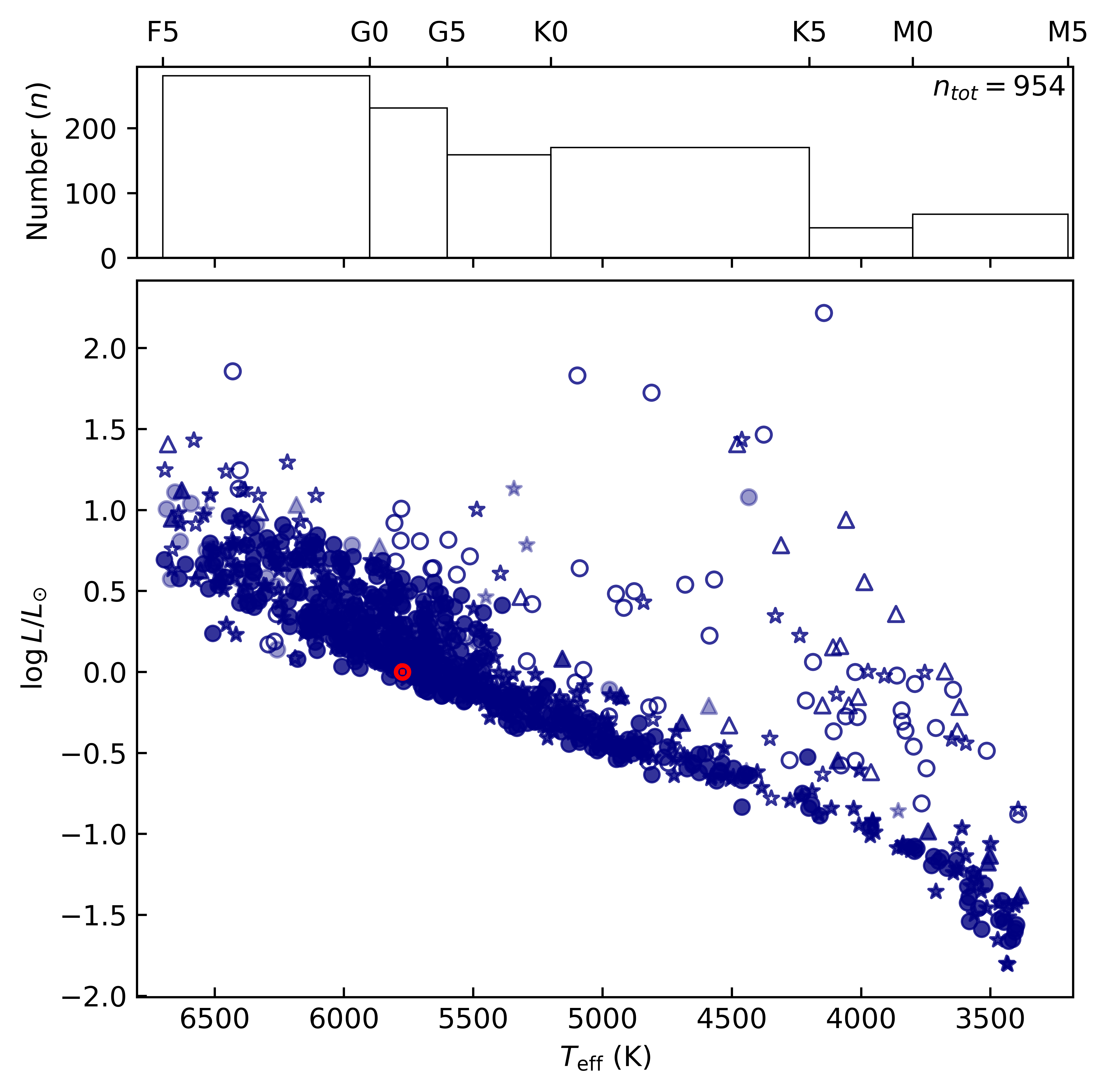}
    \caption{Bottom: HR diagram of the sample of PolarBase stars selected for our study. 
    Circles denote single objects, stars denote binary or multiple star systems where the measured $\log R'_{\rm HK}$ may be impacted by the presence of companions, and triangles indicate binary or multiple systems where the $B_l$ and $\log R^{\prime}_{\rm HK}$ are known to be impacted by blending. Filled markers are taken to be main sequence stars. Open markers have significant \ion{Li}{i} absorption lines present in their spectra, suggesting that they may be young stars. Markers with reduced opacity are stars with noisy spectra, for which it is not clear if the stars have strong \ion{Li}{i} lines. The Sun is indicated with the usual symbol in red. 
    Top: Histogram showing the breakdown of the sample by spectral classification.}
    \label{fig:HRD}
\end{figure}

\subsubsection{Rotational velocity}

The line-of-sight projected rotational velocity ($v\sin{i}$) was used to estimate the rotation period of each target (section \ref{rotation_period}). We adopted $v\sin{i}$ from \citet{Valenti2005}, \citet{Glebocki2005} or the median of published values listed in {\sc{simbad}}, as indicated in Table \ref{tab:stellar_properties}.

\subsubsection{Binary/multiple systems}

We identified binary and multiple systems using the {\sc{simbad}} database. We also inspected the LSD Stokes {\it{I}} spectral line profiles to identify any additional binary/multiple systems. The presence of multiple stellar signals in spectra and LSD profiles can result in the overestimation of the $\log R^{\prime}_{\rm HK}$ and/or $B_l$ because of blending between components. Systems for which the measured $B_l$ is impacted by the presence of companion stars are flagged in Table \ref{tab:stellar_properties}. We assume that the $\log R^{\prime}_{\rm HK}$ for all binaries/multiples may be contaminated by contributions from companions. This is because the cores of the \ion{Ca}{ii} H and K lines are intrinsically very broad compared to most metallic lines, so the contributions of companions are usually blended. 

\subsubsection{\ion{Li}{i} absorption lines to distinguish young stars from evolved subgiants}\label{sec:Li}

The presence of a strong \ion{Li}{i} line (centered at 6707.8\AA) in stellar spectra is an indicator of youth for single stars \citep{Soderblom2010}. We visually inspected the spectra to check for \ion{Li}{i} lines. For stars that were clearly located away from the main sequence, and with no/weak \ion{Li}{i} absorption lines, we assumed that the stars are  evolving off the main sequence and excluded them from our sample. For the remaining sample, we assumed stars are youthful if they have a \ion{Li}{i} line as strong as the nearby 6717.7{\AA}  \ion{Ca}{i} line.  In some cases, the spectra were noisy and we could not clearly detect \ion{Li}{i} or \ion{Ca}{i} lines, in which case we included the stars in our sample, and have indicated throughout our results that we are unsure if a strong \ion{Li}{i} line is present. Although this is a rudimentary approach to classifying stars as `youthful', we consider it to be sufficient for this study.

\onecolumn
\begin{sidewaystable}
    \centering
    \caption{Stellar properties for the PolarBase stars selected for this study. Columns (left to right) show the main identifier for each star, its GAIA ID, spectral type and luminosity classification according to {\sc{simbad}}, whether the presence of binary/multiple companion stars impact on our observations, whether significant \ion{Li}{i} absorption lines were detected in spectra, stellar effective temperature ($T_\mathrm{eff}$), luminosity ($\log \frac{L}{L_{\odot}}$), color index (B-V), surface gravity ($\log g$), radius (Rad) and projected rotational velocity ($v\sin{i}$). Asterisks ($\ast$) denote binaries/multiples for which our $B_l$ measurements are not likely to be impacted by the presence of companions, and crosses $\dagger$ denote stars where the $B_l$ measurement is impacted by companions. We assume that the measured $\log R^{\prime}_{\rm HK}$ for all binary/multiple stars may be impacted by the presence of companions. The full table is available online. }
\begin{tabular}{lcccccccccc}
\toprule
Main ID & GAIA ID  & Spectral Type & Binary impact & \ion{Li}{i} present & $T_\textrm{eff} (K) $ & $\log ({L/L_{\odot}}$) & B-V & $\log{g}$ (cm s$\mathbf{^{-2}}$) & Rad ($R_{\odot}$) & $v\sin{i}$ (kms$^{-1}$) \\ \toprule
        54 Psc   &  2802397960855105920  &       K0.5V  &             &       &  5107  & -0.266  &  0.823  &  4.37  &  0.94  &   1.1  $ ^1$  \\
      61 Cyg A   &  1872046574983507456  &         K5V  &     $\ast$  &       &  4275  & -0.795  &  1.164  &  4.43  &  0.73  &   2.0  $ ^2$  \\
        61 UMa   &  4025850731201819392  &         G8V  &     $\ast$  &       &  5458  & -0.198  &  0.704  &  4.49  &  0.89  &   2.4  $ ^1$  \\
   $\chi^1$ Ori  &  3399063235753901952  &         G0V  &     $\ast$  &       &  5854  &  0.084  &  0.557  &  4.38  &  1.07  &   9.8  $ ^1$  \\
 $\epsilon$ Eri  &  5164707970261630080  &         K2V  &     $\ast$  &       &  4871  & -0.420  &  0.910  &  4.39  &  0.87  &   2.4  $ ^1$  \\
 $\kappa^1$ Cet  &  3269362645115584640  &         G5V  &     $\ast$  &       &  5630  & -0.050  &  0.632  &  4.41  &  0.99  &   5.2  $ ^1$  \\
    $\xi$ Boo A  &  1237090738916392704  &        G7Ve  &     $\ast$  &       &  5420  & -0.238  &  0.713  &  4.51  &  0.86  &   4.6  $ ^1$  \\
   $\rho^1$ Cnc  &   704967037090946688  &      K0IV-V  &     $\ast$  &       &  5133  & -0.193  &  0.813  &  4.31  &  1.01  &   2.5  $ ^1$  \\
     $\tau$ Boo  &  1244571953471006720  &      F7IV-V  &     $\ast$  &       &  6301  &  0.529  &  0.434  &  4.17  &  1.54  &  15.0  $ ^1$  \\
 $\upsilon$ And  &   348020448377061376  &         F9V  &     $\ast$  &       &  6061  &  0.555  &  0.492  &  4.08  &  1.72  &   9.6  $ ^1$  \\
        EMSR 9   &  6049153921054413440  &         K5e  &             &  yes  &  3861  & -0.023  &  1.278  &  3.48  &  2.18  &  14.2  $ ^2$  \\
      HD 166435  &  4590489981163831296  &        G1IV  &             &       &  5728  & -0.021  &  0.601  &  4.43  &  0.99  &   7.9  $ ^1$  \\
      HD 189733  &  1827242816201846144  &         K2V  &     $\ast$  &       &  4907  & -0.462  &  0.898  &  4.46  &  0.81  &   2.9  $ ^2$  \\
      HD 190771  &  2061876742143320064  &         G2V  &     $\ast$  &       &  5680  &  0.026  &  0.617  &  4.36  &  1.06  &   4.3  $ ^1$  \\
      HD 206860  &  1772187382746856320  &        G0V+  &     $\ast$  &  yes  &  5862  &  0.054  &  0.556  &  4.41  &  1.03  &  10.6  $ ^1$  \\
      HD 224085  &  2855095251072482432  &  K2+IVeFe-1  &     $\ast$  &  yes  &  4331  &  0.345  &  1.184  &  3.52  &  2.64  &  23.0  $ ^2$  \\
      HD 75332   &   716109930305855360  &        F7Vs  &             &       &  6148  &  0.310  &  0.467  &  4.28  &  1.26  &   9.0  $ ^1$  \\
      HD 76151   &  5760701787150565888  &         G2V  &             &       &  5663  & -0.013  &  0.620  &  4.38  &  1.02  &   1.2  $ ^1$  \\
      HD 78366   &   714116137767540096  &      G0IV-V  &             &       &  5904  &  0.103  &  0.542  &  4.38  &  1.08  &   3.9  $ ^1$  \\
        HD 9986  &  2585856120791459584  &         G2V  &             &       &  5689  &  0.037  &  0.610  &  4.35  &  1.07  &   2.6  $ ^1$  \\
         BP Tau  &   164832740220756608  &      K5/7Ve  &             &  yes  &  4014  & -0.279  &  1.260  &  3.95  &  1.50  &  10.9  $ ^2$  \\
         EK Dra  &  1668690628102524672  &       G1.5V  &     $\ast$  &  yes  &  5528  & -0.054  &  0.671  &  4.39  &  1.02  &  16.8  $ ^1$  \\
         TW Hya  &  5401795662560500352  &        K6Ve  &             &  yes  &  4022  & -0.549  &  1.248  &  4.26  &  1.09  &  14.0  $ ^2$  \\
       V410 Tau  &   164518589131083136  &        K3Ve  &  $\dagger$  &  yes  &  4108  &  0.151  &  1.267  &  3.52  &  2.35  &  72.7  $ ^2$  \\
       V471 Tau  &    43789772861265792  &      K2V+DA  &     $\ast$  &  yes  &  4946  & -0.415  &  0.882  &  4.43  &  0.84  &               \\
       V830 Tau  &   147831571737487488  &        M0-1  &             &  yes  &  3841  & -0.308  &  1.256  &  3.80  &  1.58  &  33.3  $ ^3$  \\
       V889 Her  &  4524317420039164544  &         G2V  &             &  yes  &  5581  &  0.027  &  0.647  &  4.32  &  1.10  &  38.0  $ ^2$  \\
   $\pi^1$ UMa   &  1092545710514654464  &       G0.5V  &             &       &  5763  & -0.009  &  0.593  &  4.44  &  0.99  &   9.3  $ ^2$  \\
   ..... & & & &&&&& \\ \bottomrule
    \end{tabular}
    \label{tab:stellar_properties}\\
    \raggedright
    $^1$ \citet{Valenti2005}; $^2$ \citet{Glebocki2005}; $^3$ {\sc{simbad}}.
\end{sidewaystable}
\twocolumn

\section{Data}

\subsection{Spectropolarimetry with ESPaDOnS and NARVAL}
The ESPaDOnS and NARVAL spectropolarimeters have been gathering observations since 2005 and 2006 respectively. PolarBase includes observations that continue until mid-2019, prior to the commissioning of the NeoNARVAL spectropolarimeter at the TBL. Both instruments have a spectral coverage of 3700 to 10480\,{\AA}  with a resolution of $\sim$65000 in polarimetric mode. When operating in non-polarimetric mode, the instruments provide high-resolution Stokes {\it{I}} (intensity) spectra. When in polarimetric mode, they can also provide Stokes {\it{V}} (circularly polarized) spectra or Stokes {\it{Q}} and {\it{U}} (linearly polarized) spectra. Stokes {\it{V}} generally have stronger Zeeman polarization signatures compared to Stokes {\it{Q}} and {\it{U}} spectra for late-type stars, and are therefore more commonly collected for PolarBase targets. Stokes {\it{V}} spectra are obtained from a series of 4 individual `sub-exposures', each of which measures flux in two orthogonal polarization states, {\it{I}} + {\it{V}} and {\it{I}} - {\it{V}}. The positions of the orthogonally polarized beams are switched between sub-exposures by rotating the retarding rhombs of the polarimeter. A Stokes {\it{I}} spectrum can be extracted from each raw sub-exposure, and the polarized spectrum is derived from the full series of 4 sub-exposures, as described in section \ref{sec:Data reduction and calibration}. 

\subsection{Data reduction and calibration}\label{sec:Data reduction and calibration}

PolarBase provides ESPaDOnS and NARVAL observations in their reduced and calibrated format. The automatic reduction software package {\sc{libre-esprit}}, based on {\sc{esprit}} \citep[`Echelle Spectra Reduction: an Interactive Tool’,][]{Donati1997}, is applied to reduce and calibrate each sub-exposure using bias and flat-field exposures taken during the observing runs. A first wavelength calibration is carried out using a Thorium-Argon arc-lamp exposure taken on the observing night, and is then refined using telluric lines as radial velocity (RV) references. This results in a final RV accuracy of $\sim 20$ to 30 $\textrm{m s}^{-1}$ \citep{Moutou2007}. Stokes {\it{I}} spectra are extracted from each sub-exposure, continuum normalized, and are made available on PolarBase at this point.  Stokes {\it{V}} spectra are extracted from the series' of calibrated sub-exposures by dividing sub-exposures with orthogonal polarization states, which removes instrumental polarization signals \citep{Donati1997}. Stokes {\it{V}} observations available on PolarBase also include a `mean' Stokes {\it{I}} spectrum, which is derived by adding the series of 4 Stokes {\it{I}} sub-exposures together, and a `null' spectrum which is calculated by dividing spectra with identical polarization states, giving a measure of instrumental polarization (noise) and indicating the reliability of the polarimetric measurement. 

\section{Analysis}

\subsection{Least-squares deconvolution}\label{sec:LSD}

Least-squares deconvolution \citep[LSD,][]{Donati1997,Kochukhov2010} is a technique that uses the thousands of lines within a spectrum to derive a single, `mean' spectral line profile with an improved SNR. LSD assumes that most spectral lines have almost identical shapes, including identical starspot--induced intensity contrasts in Stokes {\it{I}} spectra, and polarization signatures in Stokes {\it{V}} spectra. The entire spectrum is taken to be a convolution of the mean Stokes {\it{I}} or {\it{V}} line profile, and a line pattern function that describes the locations and relative strengths of the individual spectral lines. The line pattern function is modelled using a synthetic line mask for a quiet star having similar properties to the target, and each line in the mask is weighted by a factor $w$:

\begin{equation}\label{eq:lineweight_I}
    w_{j}=\frac{d_{j}}{d_{0}} \textrm{for Stokes {\it{I}} data, or}
\end{equation} 

\begin{equation}\label{eq:lineweight_V}
    w_{j}=\frac{g_{j}\lambda_{j} d_{j}}{g_{0}\lambda_{0} d_{0}} {\textrm{for Stokes {\it{V}} data}}
\end{equation} 

where $g_j$, $\lambda_j$ and $d_j$ are the $j^{th}$ line's effective Land\'e factor, central wavelength and central depth.  The factors $g_0$, $\lambda_0$ and $d_0$ are the normalization parameters from Table \ref{tab:LSD_line_weights} \citep[an extension of Table 4 from ][]{marsden2014}. The line pattern function can thus be deconvolved from the observed spectrum to determine the mean Stokes {\it{I}} or {\it{V}} profile, as described in \citet{Donati1997} and \citet{Kochukhov2010}.

\begin{figure}
    \centering
    \includegraphics[width=0.85\columnwidth]{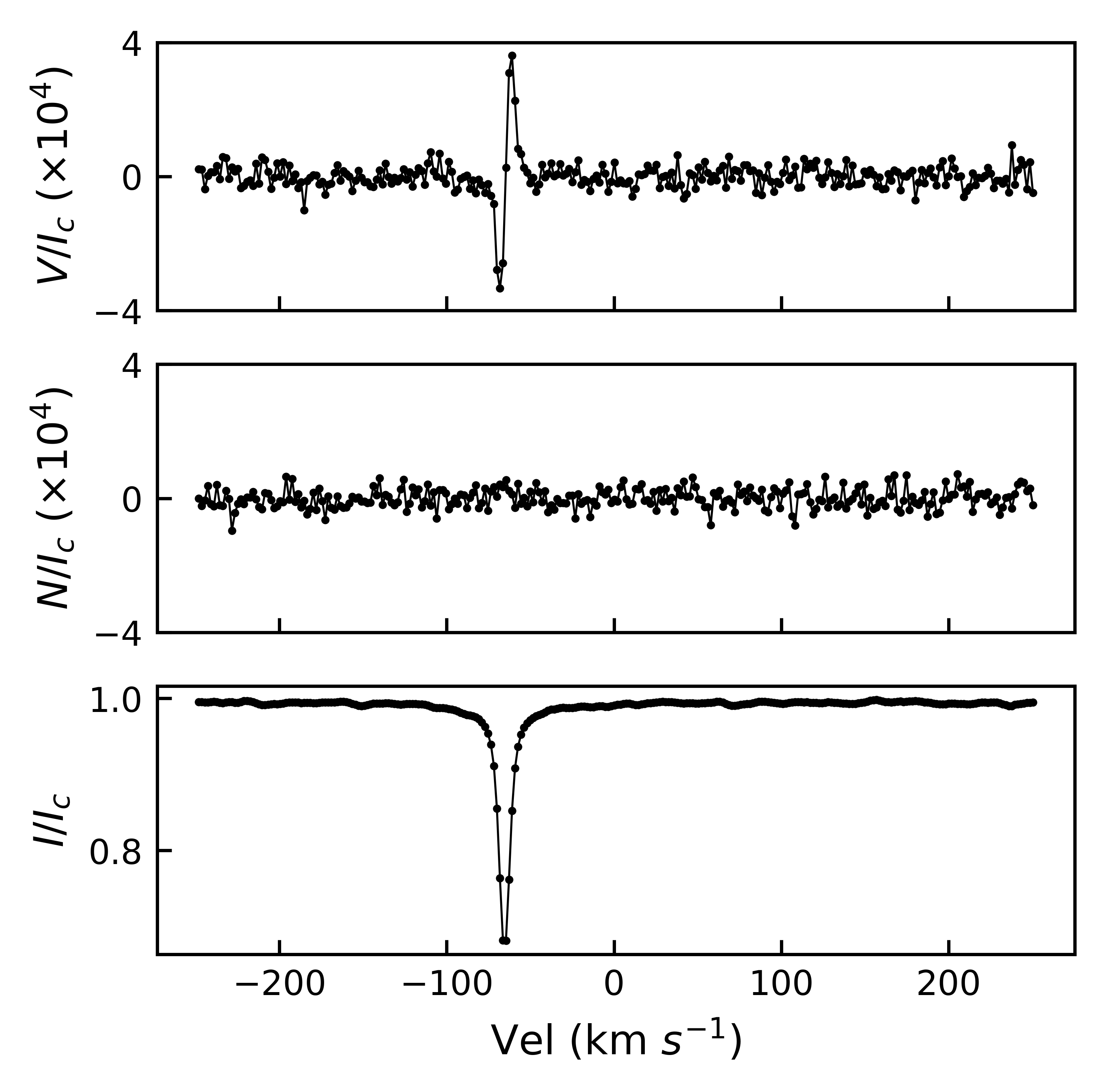}
    \caption{Example LSD profile for 61 Cygni A. Axes (top to bottom) show the mean Stokes {\it{V}} profile, Null profile and mean Stokes {\it{I}} profile. Note that the y-axis scale has been enlarged for the Stokes {\it{V}} and Null profiles for clarity.}
    \label{fig:LSDplot}
\end{figure}

We performed LSD using the automated technique of \citet{Donati1997}. We generated synthetic spectral line masks using stellar atmospheric models from the Vienna Atomic Line Database \citep[VALD,][]{Kupka2000}, for stellar temperatures ranging between 3000 and 6500\,K in steps of 500\,K, and $\log g$ between 2.0 and 5.0\,cm\,s$^{-2}$ in steps of 0.5\,cm\,s$^{-2}$. Line masks excluded lines with depth $\leq$ 10 per cent of the continuum. For each target we selected the line mask with the nearest $T_\mathrm{eff}$ and $\log g$, and used the line normalization parameters from Table \ref{tab:LSD_line_weights}.  An example LSD line profile is shown in Figure \ref{fig:LSDplot},  and further details of the LSD technique can be found in \citet{Donati1997} and \citet{Kochukhov2010}.

\begin{table}
    \centering
    \caption{Table of normalization parameters used for LSD \citep[extension of Table 4 from ][]{marsden2014}.}
    \begin{tabular}{llll}
    \toprule
$\mathbf{T_{eff}}$ (K) & ${g_0}$ & ${d_0}$ & ${\lambda_0}$ \\ \toprule
3000	&	1.22	&	0.55	&	690 \\
3500	&	1.22	&	0.55	&	670 \\
4000	&	1.22	&	0.55	&	650\\
4500	&	1.22	&	0.55	&	630\\
5000	&	1.22	&	0.54	&	610\\
5500	&	1.22	&	0.53	&	590\\
6000	&	1.22	&	0.51	&	570\\
6500	&	1.21	&	0.49	&	560\\
\bottomrule
    \end{tabular}
    \label{tab:LSD_line_weights}
\end{table}

\subsection{Radial velocity}\label{sec:RV}

We used the LSD line profiles to measure the precise radial velocity of each target during each observation. We fitted a pseudo-Voigt profile (convolution of a Gaussian and a Lorentzian) to each Stokes {\it{I}} LSD profile and took the central velocity of the fitted model as the RV. We then RV-corrected each LSD profile and the corresponding spectra. In Table \ref{tab:results} we show the error-weighted average RV and standard deviation for each target. Note that the particularly large standard deviations shown in Table \ref{tab:results}, such as for V471 Tau, are generally related to the changes in the RV throughout binary orbits. 
\subsection{Chromospheric activity and variability}

\subsubsection{Chromospheric S-index}\label{sec:S-index}

The S-index \citep{Vaughan1978} is the ratio of the fluxes in the \ion{Ca}{ii} H and K lines to the continuum flux either side of the H and K lines. It is calculated as 
\begin{equation}\label{eq:Sindex}
    S_{MW}=\frac{a F_H + b F_K}{c F_{R_{\rm HK}} + d F_{V_{\rm HK}}} +e ,
\end{equation}
where $F_{H}$ and $F_{K}$ are the fluxes in two triangular bandpasses centred on the cores of the \ion{Ca}{ii} H and K lines (3968.469\,{\AA} and 3933.663\,{\AA})  with widths of 2.18\,{\AA} (at the base), and $F_{R_{\rm HK}}$ and $F_{V_{\rm HK}}$ are the fluxes in two rectangular 20\,{\AA} bandpasses centered on the continuum either side of the H and K lines at 3901.07\,{\AA} and 4001.07\,{\AA}. The coefficients a, b, c, d and e are derived in \cite{marsden2014} to convert S-indices from NARVAL and ESPaDOnS to the Mount Wilson S-index scale. 

We calculated S-indices using individual Stokes {\it{I}} sub-exposures as these were available on PolarBase for a large majority of our targets, as opposed to `mean' Stokes {\it{I}} profiles which were only available for stars with full series' of 4 sub-exposures. S-index errors were calculated by propagating the uncertainties computed during the reduction process for each spectral bin of the normalized Stokes {\it{I}} spectra through Equation \ref{eq:Sindex}.

 For targets with $\geq10$ observations we removed outliers in the S-indices by excluding values outside 3 standard deviations from the mean \citep[as in][]{Gomes2020}. For smaller data sets we found this to inappropriately exclude values that were not outliers, so we did no $\sigma$ clipping for these stars. Some observations measured intensities below zero in the region of the \ion{Ca}{ii} H\&K lines. We filtered out observations where >1\% of pixels within the H and K bandpasses had intensities < 0. 
 
 Table \ref{tab:results} shows the mean S-index for each target ($\langle S \rangle$), weighted by the errors in individual S-index observations. We propagated S-index errors through the weighted mean equation, and they are so small that we have excluded them from Table \ref{tab:results}. We have also measured the peak-to-peak S-index range ($\Delta S$) as a representation of the amplitude of activity variability, as in \citet{Saar2002}. We found the peak-to-peak amplitude to be a more realistic representation of activity variability compared to the standard deviation, since chromospheric activity cycles are not strictly periodic nor symmetrical about the mean.

\subsubsection{Conversion from S-index to $R^{\prime}_{\rm HK}$}\label{sec:Rhk}

Chromospheric \ion{Ca}{ii} H\&K emissions include a basal flux contribution \citep{Schrijver1987}, unrelated to magnetic activity, and a photospheric contribution due to magnetic heating \citep{Noyes1984}. Both can be estimated based on the B-V colour of a star \citep{Middelkoop1982,Rutten1984}, and must be removed when comparing the chromospheric activity of stars across the HR diagram, to obtain a purely chromospheric activity index, $R'_{\rm HK}$. This is often done using the method of \citet[][equation \ref{eq:R'hk}]{Noyes1984},

\begin{equation}\label{eq:R'hk}
    R'_{\rm HK}= 1.34 \times 10^{-4} \cdot C_{cf} \cdot S_{MW} - R_{phot},
\end{equation}
where $C_{cf}$ is the bolometric correction factor and $R_{phot}$ is the photospheric contribution to the H and K bandpasses. However, the original calibration of the photospheric correction is valid only for $0.44 \leq B-V \leq 0.82$, while our sample cover $0.37 \leq B-V \leq 1.4$. Others have calibrated the \citet{Noyes1984} photospheric and bolometric corrections to a larger spectral range, such as \citet{Suarez_Mascareno2015} ($0.4 \leq B-V \leq 1.9$), \citet{Astudillo2017} ($0.54 \leq B-V \leq 1.9$) and \citet{BoroSaikia2018}. We used the \citet{Suarez_Mascareno2015} conversion, where $C_{cf}$ and $R_{phot}$ are calculated as

\begin{equation}\label{eq:Ccf}
    \log_{10}{C_{cf}} = 0.668(B-V)^3 - 1.270(B-V)^2 - 0.645(B-V) - 0.443
\end{equation}

\begin{equation}\label{eq:Rphot}
    \log_{10}{R_{phot}} = 1.48 \times 10^{-4} \cdot \exp[{-4.3658(B-V)}].
\end{equation}

Three stars within our sample are slightly outside the calibrated B-V range of the \citet{Suarez_Mascareno2015} conversion, with B-V between 0.37 and 0.40. We still used the \citet{Suarez_Mascareno2015} conversion for these stars, and we do not expect that this has any significant impact on the key findings of this study. We calculated the mean $\log R'_{\rm HK}$ and $\log R'_{\rm HK}$ range ($\log \Delta R'_{\rm HK}$) based on the mean S-index and S-index range from Table \ref{tab:results}, and the B-V from Table \ref{tab:stellar_properties}. 

\subsection{Longitudinal magnetic field}\label{sec:B_l}

The longitudinal magnetic field, $B_l$, is the line-of-sight magnetic field averaged over the entire stellar disk. For the stars within our sample that have Stokes {\it{V}} profiles available in PolarBase, we measured $B_l$ using 

\begin{equation}\label{eq:bl}
    B_{l}=-2.14\times{10^{11}} \frac{\int{vV(v) dv}}{\lambda g c \int{[1-I(v)]dv}}
\end{equation}
from \citet{Donati1997}, where $B_l$ is in gauss, {\it{V(v)}} and {\it{I(v)}} are the RV-corrected Stokes {\it{V}} and {\it{I}} LSD profiles respectively, which are normalised by the unpolarized continuum ($I_c$), $\lambda$ and $g$ are the normalization parameters from Table \ref{tab:LSD_line_weights}, c is the speed of light in km\,s$^{-1}$ and $v$ is velocity in km\,s$^{-1}$. 

The velocity domain over which to integrate is unique for each target, and was chosen to include the entire Stokes {\it{V}} polarization signal while minimizing noise. We used the mean full-width at half-maximum (FWHM) of the Stokes {\it{I}} LSD profile (Table \ref{tab:results}) to determine an appropriate domain according to the equation
\begin{equation}\label{eq:Bl_domain}
    v\in (\pm\,\alpha FWHM) ,
\end{equation}
where $\alpha=1.3$ for FWHM < 40 $\textrm{km s}^{-1}$. For a small selection of stars with very broad profiles (FWHM $\geq$ 40 $\textrm{km s}^{-1}$) $\alpha=0.8$ was more appropriate.  

The error for each $B_l$ observation was determined by propagating the uncertainties computed during the reduction process for each spectral bin of the normalized spectrum through equation \ref{eq:bl}. To determine the mean and range of the $B_l$, we first filtered out observations with very large uncertainties (>200G), and those for which the uncertainty in the $B_l$ was greater than the measured value. For stars with >10 observations, we also removed $B_l$ values that were more than 3 standard deviations from the mean. We  measured the $B_l$ from the Null profile, $B_{l, N}$, using the same method as above. Non-zero values of $B_{l, N}$ that are significant with respect to their uncertainties, and significant with respect to the $B_l$ measured from the Stokes {\it{V}} profile, may indicate that our magnetic field detection is spurious. Therefore, we also filtered out observations with large detections in the Null spectrum ($|B_{l, N}|\geq |B_l|$). Table \ref{tab:results} shows the error-weighted mean, unsigned longitudinal field strength from the Stokes {\it{V}} and Null profiles ($\langle |B_l| \rangle$ and $\langle |B_{l, N}|\rangle$ respectively), as well as the peak-to-peak $B_l$ range ($\Delta B_l$).  Uncertainties in the $\langle |B_l| \rangle$ and $\langle |B_{l, N}| \rangle$ were determined by propagating the errors in individual measurements through the weighted mean equation. We also show the mean uncertainty in the $B_{l, N}$ values, $\mathbf{\langle |\sigma_{B_{l, N}}| \rangle}$, which provides an indication of the significance of the average detection in the Null spectrum. There are a few stars with large $\langle|B_{l, N}|\rangle$ values, but these are not significant relative to their average uncertainties, $\mathbf{\langle |\sigma_{B_{l, N}}| \rangle}$. 

\subsection{Stellar rotation}\label{rotation_period}

We estimated stellar rotation periods ($P_{\mathrm{est,} i=60\degr}$) by assuming solid-body rotation, taking a constant inclination angle for all stars of $60\degr$, and using the stellar radius and $v\sin{i}$ from Table \ref{tab:stellar_properties}. 

\section{Results and discussion}\label{sec:results}

\begin{figure*}
    \centering
    \includegraphics[width=1\linewidth]{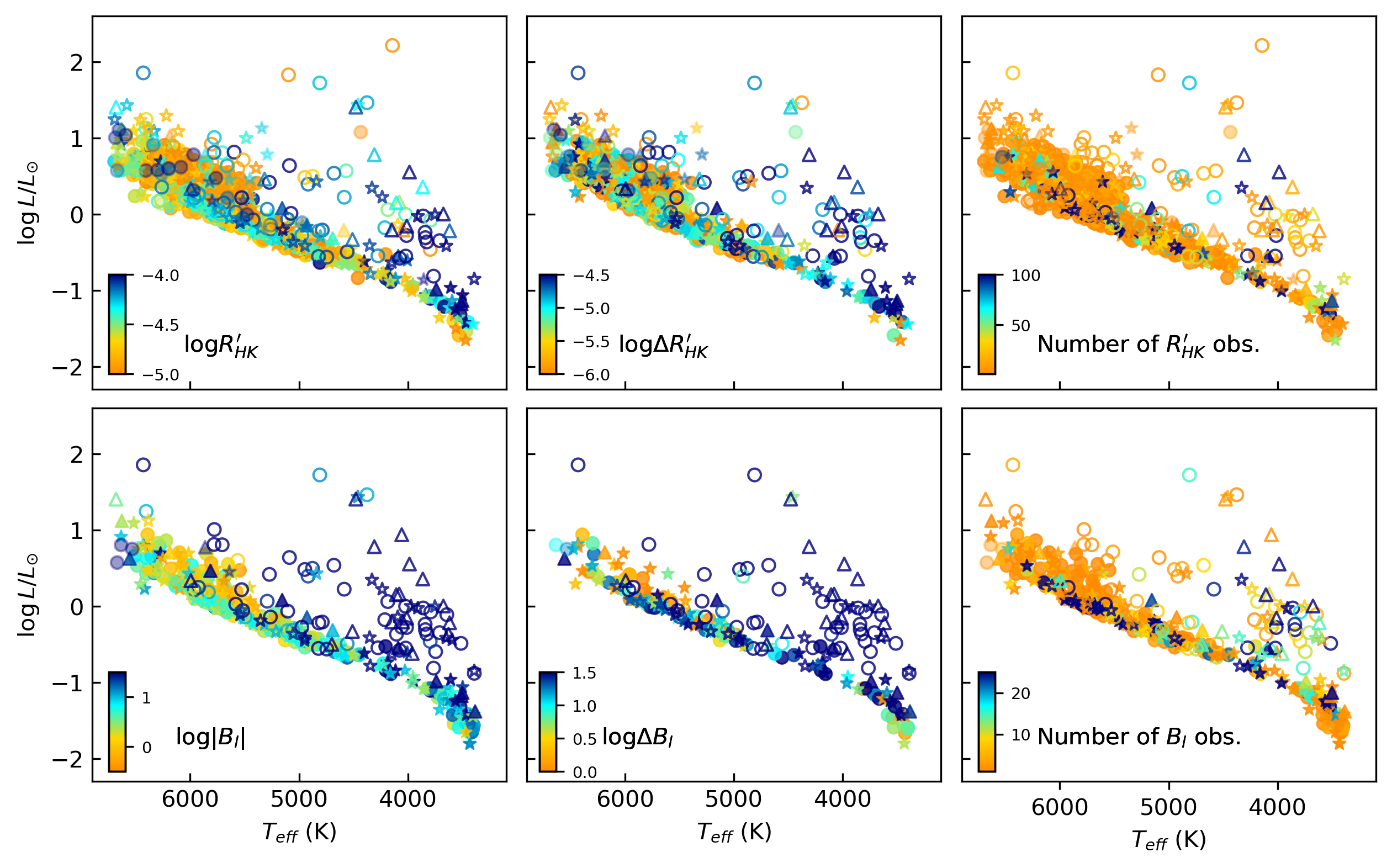}
    \caption{Top: HR diagrams showing (left to right) mean chromospheric activity ($\log R^{\prime}_{\rm HK}$), amplitude of chromospheric activity variability ($\log \Delta R^{\prime}_{\rm HK}$) and number of observations used to measure the mean $\log R^{\prime}_{\rm HK}$. 
    Bottom: HR diagrams showing (left to right) the mean, unsigned surface--averaged longitudinal magnetic field strength ($\log|B_l|$), amplitude of magnetic field strength variability ($\log \Delta B_l$) and number of observations used to measure the mean $|B_l|$.  For all plots, the marker shapes and fill-styles are the same as in Figure \ref{fig:HRD}.  \ion{Li}{i} abundant stars, which we take to be young stars, are indicated by open markers, while mature stars are shown by filled markers. Stars for which we were unable to determine if significant \ion{Li}{i} lines were present are shown by filled markers with a reduced opacity. Circles denote single objects, while star symbols indicate binaries for which $B_l$ measurements do not appear to be impacted by blending between companions, and triangles represent binaries for which $B_l$ is known to be impacted. }
    \label{fig:HRDresults}
\end{figure*}

We measured chromospheric $\log R^{\prime}_{\rm HK}$ indices for 913 stars. For 621 of these we had multiple $\log R^{\prime}_{\rm HK}$ observations and could estimate the amplitude of the chromopsheric activity variability, $\log \Delta R^{\prime}_{\rm HK}$. We measured the mean, unsigned, surface-averaged magnetic field strength, $\langle |B_l| \rangle$, for 425 stars, and estimated the $B_l$ variability amplitude for 233 of this sample. Stars for which we were able to measure both the $\log R^{\prime}_{\rm HK}$ and $B_l$ totalled 383. Table \ref{tab:results} provides the measured parameters (the full table is available online only), which are also shown on the HR diagram in Figure \ref{fig:HRDresults}.

\begin{sidewaystable*}
    \centering
    \caption{ Activity and magnetic field results for the stars shown in Table \ref{tab:stellar_properties}. The full table is available online. Columns show the weighted mean RV measured from the LSD profiles for each star, where the uncertainty in the RV is its standard deviation across observations, the mean LSD full-width at half-maximum (FWHM), the number of S-index observations ($N_S$), error-weighted mean and peak-to-peak range of the S-indices ($\langle S \rangle$ and $\Delta S$), which are also converted to an ${R'_{\rm HK}}$ mean and range ($\log \langle R'_{\rm HK} \rangle$ and $\log \Delta R'_{\rm HK}$). Also shown are the number of $B_l$ observations ($N_{B_l}$), the velocity range used to measure $B_l$, the mean unsigned $B_l$ ($\langle |B_{l}| \rangle$) and peak-to-peak $B_l$ range ($\Delta B_{l}$), mean strength of the magnetic field in the Null profile ($\langle |B_{l, N}| \rangle$), mean error on the $B_{l, N}$ measurements ($\langle |\sigma_{B_{l, N}}| \rangle$) and estimated stellar rotation period, $P_{\mathrm{est,} i=60\degr}$.}
   \begin{tabular}
   {lccccccccccccccccc}
\toprule
 \multirow{2}{*}{ID} & \multirow{2}{*}{RV ($\textrm{km s}^{-1}$)} & LSD FWHM  &  \multirow{2}{*}{$N_S$} & \multirow{2}{*}{$\langle S \rangle$} & \multirow{2}{*}{$\Delta S$} & \multirow{2}{*}{$\log{\langle R^{\prime}_{\rm HK}} \rangle$} & \multirow{2}{*}{$\log \Delta {R^{\prime}_{\rm HK}}$} & \multirow{2}{*}{$N_{B_l}$} & Vel. range & {$\langle |B_{l}| \rangle$} & {$\Delta B_{l}$}  & {$\langle |B_{l, N}| \rangle$}
 & {$\langle |\sigma_{B_{l, N}}| \rangle$}
 & {$P_{\mathrm{est,} i=60\degr}$} \\
 && ($\textrm{km s}^{-1}$) &&&&&&&($\textrm{km s}^{-1}$)&(G)&(G)&(G)&(G)&(d)
 \\\toprule
         54 Psc   &   -32.7 $\pm$ 0.1    &    7.8  &   273  &   0.187  &   0.079  &  -4.914  &  -5.162  &   61  &  -43:-23  &    2.1 $\pm$ {0.1}  &    8.8  &   0.3 $\pm$ {0.1}  &   0.4  &  37.4  \\
      61 Cyg A   &   -65.5 $\pm$ 0.4    &    7.7  &   720  &   0.637  &   0.326  &  -4.725  &  -4.995  &  157  &  -76:-56  &    5.4 $\pm$ {0.0}  &   25.6  &   0.3 $\pm$ {0.0}  &   0.4  &  16.0  \\
        61 UMa   &    -5.4 $\pm$ 0.1    &    8.1  &   323  &   0.313  &   0.107  &  -4.529  &  -4.903  &   69  &    -15:5  &    2.7 $\pm$ {0.1}  &   22.9  &   0.3 $\pm$ {0.1}  &   0.4  &  16.2  \\
   $\chi^1$ Ori  &   -13.8 $\pm$ 3.2    &   14.7  &   559  &   0.326  &   0.088  &  -4.397  &  -4.846  &  101  &    -31:7  &    3.1 $\pm$ {0.1}  &   16.8  &   0.7 $\pm$ {0.1}  &   0.9  &   4.8  \\
 $\epsilon$ Eri  &    16.5 $\pm$ 0.5    &    8.2  &   664  &   0.498  &   0.167  &  -4.500  &  -4.938  &  155  &     6:26  &    4.0 $\pm$ {0.0}  &   25.1  &   0.2 $\pm$ {0.0}  &   0.3  &  15.8  \\
 $\kappa^1$ Cet  &    19.2 $\pm$ 0.1    &   10.3  &   298  &   0.351  &   0.112  &  -4.410  &  -4.812  &   67  &     6:32  &    4.6 $\pm$ {0.1}  &   22.3  &   0.6 $\pm$ {0.1}  &   0.7  &   8.4  \\
    $\xi$ Boo A  &     2.0 $\pm$ 0.3    &    9.7  &   675  &   0.452  &   0.109  &  -4.347  &  -4.905  &  167  &   -10:14  &    7.2 $\pm$ {0.0}  &   30.8  &   0.4 $\pm$ {0.0}  &   0.6  &   8.2  \\
   $\rho^1$ Cnc  &    27.5 $\pm$ 0.3    &    7.9  &   399  &   0.201  &   0.128  &  -4.865  &  -4.943  &   45  &    17:37  &    2.2 $\pm$ {0.0}  &   11.7  &   0.2 $\pm$ {0.0}  &   0.3  &  17.8  \\
     $\tau$ Boo  &   -16.3 $\pm$ 0.7    &   22.1  &  1634  &   0.190  &   0.099  &  -4.737  &  -4.675  &  142  &   -44:12  &    2.0 $\pm$ {0.1}  &   14.6  &   0.9 $\pm$ {0.1}  &   1.0  &   4.5  \\
 $\upsilon$ And  &   -28.4 $\pm$ 0.1    &   15.2  &   486  &   0.152  &   0.020  &  -4.951  &  -5.421  &    1  &  -48:-10  &    0.8 $\pm$ {0.5}  &      -  &   0.4 $\pm$ {0.5}  &   0.5  &   7.8  \\
        EMSR 9   &    -6.8 $\pm$ 1.6    &   22.5  &     6  &   5.819  &   4.122  &  -3.936  &  -4.084  &  138  &   -36:22  &  131.3 $\pm$ {1.2}  &  558.3  &  11.3 $\pm$ {1.2}  &  14.3  &   6.7  \\
      HD 166435  &   -14.2 $\pm$ 0.1    &   13.1  &   290  &   0.393  &   0.068  &  -4.326  &  -4.997  &   65  &    -29:3  &    7.1 $\pm$ {0.1}  &   29.7  &   0.7 $\pm$ {0.1}  &   1.0  &   5.5  \\
      HD 189733  &    -2.1 $\pm$ 0.3    &    8.9  &   583  &   0.541  &   0.300  &  -4.447  &  -4.669  &  133  &    -14:8  &    4.5 $\pm$ {0.0}  &   26.2  &   0.4 $\pm$ {0.0}  &   0.6  &  12.3  \\
      HD 190771  &   -25.8 $\pm$ 1.1    &    9.7  &   603  &   0.328  &   0.059  &  -4.435  &  -5.078  &  139  &  -38:-14  &    4.1 $\pm$ {0.1}  &   25.0  &   0.5 $\pm$ {0.1}  &   0.6  &  10.8  \\
      HD 206860  &   -16.6 $\pm$ 0.1    &   15.8  &   638  &   0.321  &   0.067  &  -4.407  &  -4.961  &  131  &    -37:3  &    6.4 $\pm$ {0.1}  &   22.7  &   1.1 $\pm$ {0.1}  &   1.6  &   4.3  \\
      HD 224085  &  -17.0 $\pm$ 56.8    &   30.7  &   483  &   2.960  &   2.010  &  -4.073  &  -4.237  &   63  &   -22:56  &   61.9 $\pm$ {0.5}  &  325.4  &   3.4 $\pm$ {0.5}  &   4.2  &   5.0  \\
      HD 75332   &     4.4 $\pm$ 0.4    &   14.6  &   416  &   0.270  &   0.038  &  -4.465  &  -5.118  &   73  &   -13:23  &    3.7 $\pm$ {0.2}  &   18.6  &   1.2 $\pm$ {0.2}  &   1.7  &   6.1  \\
      HD 76151   &    32.2 $\pm$ 0.1    &    7.7  &   271  &   0.235  &   0.056  &  -4.633  &  -5.102  &   60  &    23:43  &    2.9 $\pm$ {0.1}  &    8.9  &   0.3 $\pm$ {0.1}  &   0.4  &  37.4  \\
      HD 78366   &    26.3 $\pm$ 0.1    &    9.3  &   384  &   0.248  &   0.046  &  -4.555  &  -5.110  &   83  &    14:38  &    3.8 $\pm$ {0.1}  &   16.5  &   0.5 $\pm$ {0.1}  &   0.6  &  12.1  \\
        HD 9986  &   -20.8 $\pm$ 0.2    &    8.6  &   335  &   0.177  &   0.035  &  -4.815  &  -5.296  &   48  &  -32:-10  &    1.3 $\pm$ {0.1}  &    6.0  &   0.4 $\pm$ {0.1}  &   0.6  &  18.1  \\
         BP Tau  &    15.6 $\pm$ 1.5    &   13.4  &   158  &   8.170  &  10.245  &  -3.757  &  -3.657  &   85  &    -1:33  &  308.6 $\pm$ {1.9}  &  576.3  &  12.4 $\pm$ {1.7}  &  15.4  &   6.0  \\
         EK Dra  &   -20.3 $\pm$ 0.4    &   24.5  &   319  &   0.608  &   0.236  &  -4.163  &  -4.528  &   72  &   -51:11  &   20.8 $\pm$ {0.7}  &   96.8  &   3.8 $\pm$ {0.7}  &   6.0  &   2.7  \\
         TW Hya  &    12.7 $\pm$ 0.4    &   11.0  &   161  &  12.185  &  14.324  &  -3.562  &  -3.491  &   65  &    -1:27  &  262.2 $\pm$ {1.0}  &  552.5  &   4.6 $\pm$ {0.8}  &   6.6  &   3.4  \\
       V410 Tau  &    16.1 $\pm$ 5.5    &   98.1  &   118  &   2.208  &   1.327  &  -4.342  &  -4.557  &   84  &   -62:94  &  136.9 $\pm$ {5.0}  &  655.3  &  34.6 $\pm$ {5.0}  &  45.4  &   1.4  \\
       V471 Tau  &   1.8 $\pm$ 190.0    &  120.5  &   622  &   1.097  &   1.235  &  -4.104  &  -4.035  &   71  &   72:264  &   89.9 $\pm$ {4.6}  &  457.7  &  25.1 $\pm$ {4.6}  &  38.6  &     -  \\
       V830 Tau  &    17.2 $\pm$ 8.5    &   43.5  &     8  &   5.552  &   0.487  &  -3.918  &  -4.972  &  102  &   -17:51  &   76.9 $\pm$ {1.8}  &  145.4  &  13.6 $\pm$ {1.8}  &  17.7  &   2.1  \\
       V889 Her  &   -23.1 $\pm$ 0.7    &   53.3  &   283  &   0.613  &   0.172  &  -4.139  &  -4.640  &   72  &   -65:19  &   33.0 $\pm$ {0.9}  &  119.3  &   5.4 $\pm$ {0.9}  &   7.7  &   1.3  \\
   $\pi^1$ UMa   &   -12.6 $\pm$ 0.1    &   15.3  &   337  &   0.372  &   0.065  &  -4.349  &  -5.014  &   66  &    -31:7  &    6.1 $\pm$ {0.1}  &   27.6  &   0.9 $\pm$ {0.1}  &   1.2  &   4.7  \\
..... & & & &&&&& \\ \bottomrule
    \end{tabular}
    \label{tab:results}
\end{sidewaystable*}

The sample of stars with measured $B_l$ includes a greater number of very cool stars compared to the $\log R^{\prime}_{\rm HK}$ sample, as shown in Figure \ref{fig:HRDresults}. This is because cooler stars emit weakly in the blue spectral wavelengths, where the \ion{Ca}{ii} H\&K lines lie, so SNRs in this domain are particularly poor. The $\log R'_{\rm HK}$ sample consists of a greater number of mature, magnetically and chromospherically inactive stars compared to $B_l$ sample, due to the lower-limit for detecting the large-scale magnetic fields for inactive stars.

Many of the stars we identified as young stars are located away from the main sequence in Figure \ref{fig:HRDresults}, but several also appear to populate the main sequence. Given that those on the main sequence are highly active and have strong \ion{Li}{i} absorption lines, it is possible that they are zero-age main sequence stars. \citet{Gomes2020} similarly found that their main sequence population included a small number of highly active stars, which had uncertain isochronal ages. They also concluded that the stars were probably younger compared to others on the main sequence. It is also possible that the positions of these stars on the main sequence is a reflection of our assumption of solar-metallicity when deriving the stellar temperatures and luminosities (section \ref{sec:params_from_MIST}). 

For stars with a large number of observations (e.g. $\geq100$), the means and amplitudes we report in Table \ref{tab:results} characterize the stellar variability on a rotational time-scale, as well as modulations related to magnetic and chromospheric activity cycles. For stars with fewer observations, the reported mean levels of chromospheric activity and magnetic field strength will be relatively poorly constrained. The $\log \Delta R^{\prime}_{\rm HK}$ and $\Delta B_l$ will be underestimated for targets with few observations, and are likely to only characterise the rotational modulation of chromospheric activity and magnetic field strength.

\subsection{Mean chromospheric activities and large-scale magnetic field strengths}

Figure \ref{fig:SvBl} compares the mean $\log R^{\prime}_{\rm HK}$ with mean $\log |B_l|$ across sub-samples of F, G, K and M stars. 
The number of stars in each sub-sample is shown at the top of each plot, along with the Pearson correlation coefficient, $\rho$.  Marker colour scales with the number of observations for each star, and marker shapes/styles are the same as in Figures \ref{fig:HRD} and \ref{fig:HRDresults}.  The black--line histograms indicate the distributions of the mean $\log |B_l|$ and mean $\log R'_{\rm HK}$ across each sample of young stars, while the grey shaded histograms indicate the data distributions for mature stars and those with noisy spectra for which we could not determine if significant \ion{Li}{i} lines were present. In Figure \ref{fig:AppA} we also show histograms for our chromospheric activity data combined with the data presented by \citet{BoroSaikia2018} and \citet{Gomes2020}.

\subsubsection{Correlation between mean $\log R^{\prime}_{\rm HK}$ and $\log |B_l|$}\label{sec:gradient_change}

Chromospheric activity across our sample ranges from $\log R'_{\rm HK}\sim-3.5$ to -5.0, and magnetic field strengths range from $\log |B_{l}|\sim-0.7$ to 2.7 ($\sim0.2$ to 500G). Across all spectral types we find a positive correlation between mean $\log R^{\prime}_{\rm HK}$ and mean $\log |B_l|$, although the correlation is only marginal for F and M stars. For all spectral types, the correlation is improved when considering only stars with a high number of observations ($\geq100$ observations, indicated by dark blue markers). Binary/multiple stellar systems generally lie along the same trend as single stars.  \citet{marsden2014} also found a positive correlation between chromospheric activity and the upper envelope of the longitudinal field strength in their snapshot study of 170 F, G and K stars from the BCool sample, and \citet{Reiners2022} found an approximately linear power-law relation between nonthermal \ion{Ca}{ii} H\&K emissions and magnetic flux (measured using Zeeman broadening) for their sample of 292 M dwarfs. 

\begin{figure*}
    \centering
    \includegraphics[width=0.85\linewidth]{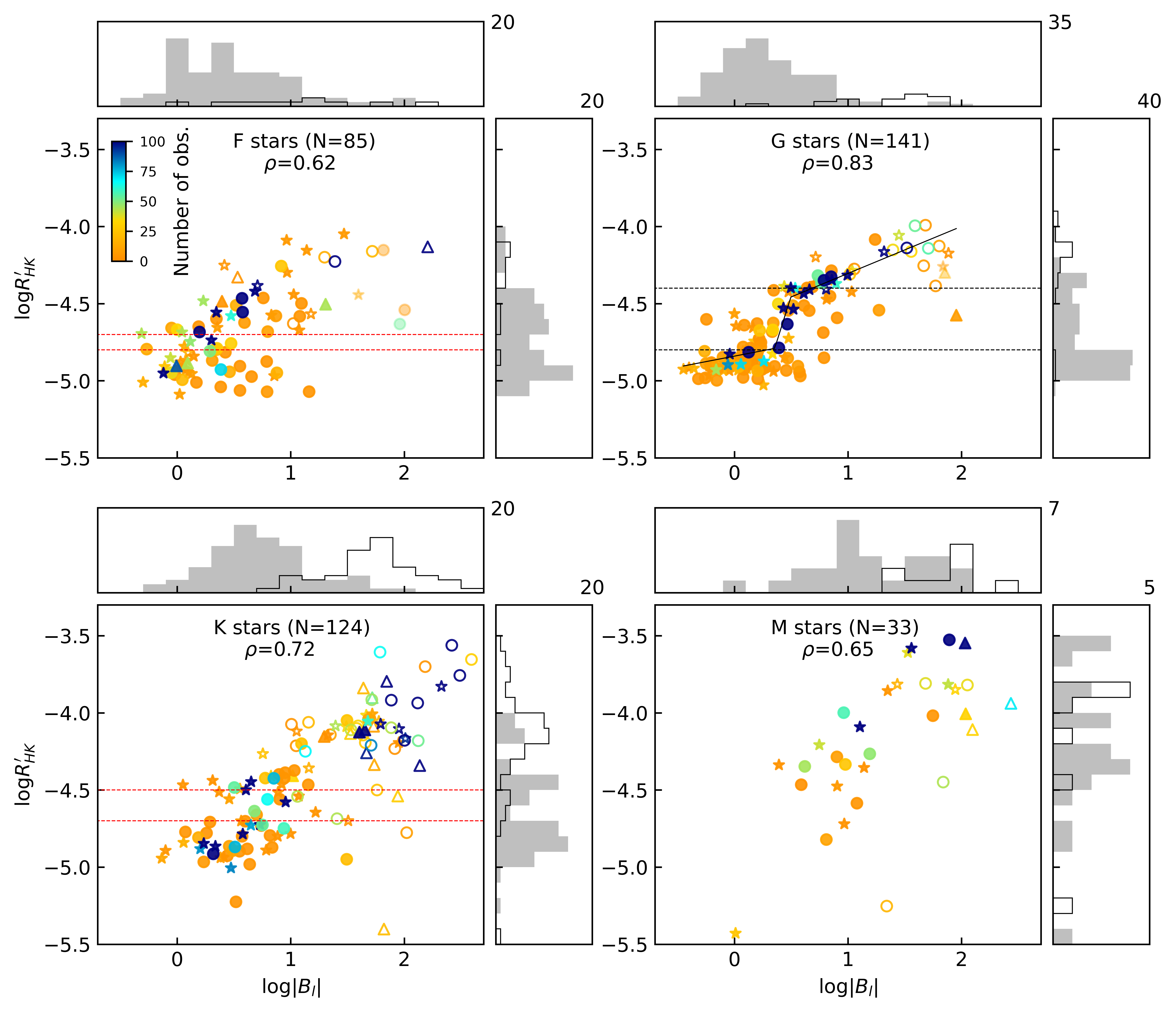}
    \caption{Mean chromospheric activity versus mean unsigned longitudinal field strength for our sample, separated by spectral type. Marker shapes and fill-styles are the same as in Figure \ref{fig:HRD}. Marker color scales with the number of observations. We take the number of observations as the average of the number of $\log R^{\prime}_{\rm HK}$ observations and $4\times B_l$ observations (since $4\times$ Stokes {\it{I}} observations are required to measure $B_l$).  The grey shaded histograms above and to the right of each plot show the distributions of the mean $\log|B_l|$ and $\log R^{\prime}_{\rm HK}$ across each sample of main sequence stars. The black line histograms indicate the distributions of data for young, lithium rich stars only. The dashed lines indicate the limits of possible reduced populations in the activity distributions; the lines are red for the F and K stars because we did not find the reduced populations to be significant with respect to the histogram uncertainties in Figure \ref{fig:AppA}. The solid black line shown for G stars indicates a continuous piece-wise linear fit to the data.  }
    \label{fig:SvBl}
\end{figure*}

Interestingly, our results suggest that the relationship between the mean $\log R^{\prime}_{\rm HK}$ and mean $\log |B_l|$ may be marked by multiple regimes. These are the most clear for the sample of G stars in Figure \ref{fig:SvBl}, possibly because this sample is the largest compared to the other stellar types. For illustrative purposes in Figure \ref{fig:SvBl} we show a continuous, piece-wise linear function (black line) which we fitted to the G-star data using a weighted least-squares fit (wherein data is weighted by the number of observations). For G stars with $\log |B_l|$ between $\sim2$ and 0.4, and $\log R'_{\rm HK}$ between $\sim-4.0$ and -4.4, chromospheric activity and magnetic field strengths decrease simultaneously. 
At $\log |B_l|\sim 0.4$, the field strength remains almost constant while $\log R'_{\rm HK}$ decreases from $\sim$-4.4 to -4.8. The relation then reverts back to decreasing $\log R'_{\rm HK}$ with decreasing $|B_l|$ for $\log |B_l|<0.4$ and $\log R'_{\rm HK}<-4.8$.

We considered if the multiple regimes could be related to some systematic issue with the estimation of the mean magnetic field strength.  For example, it is possible that the mean $|B_l|$ could be overestimated for stars with $\log R'_{\rm HK}\leq-4.4$ because we filter out observations where the measured $B_l$ is below the noise level (measured in the Null profile). Thus, for inactive stars, the mean $|B_l|$ could relate to magnetic maxima only, and this might explain why inactive stars appear to be `shifted' to the right in $|B_l|$ space. However, when we plotted the maximum $|B_l|$ against the mean $\log R'_{\rm HK}$ for the sample shown in Figure \ref{fig:SvBl}, distinct regimes were still evident. Therefore, overestimation of the mean $|B_l|$ for inactive stars is unlikely to fully explain the relationship shown in Figure \ref{fig:SvBl}.


Similarly, a systematic underestimation of the mean $|B_l|$ for low activity stars would not fully explain the observed shape of the $\log R'_{\rm HK}$ - $\log |B_l|$ relation. The $|B_l|$ could be underestimated for inactive stars due to their low rotational velocity, which may limit the spatial resolution of the Stokes {\it{V}} and {\it{I}} spectral line profiles. Or, if the magnetic field were concentrated into smaller spatial scales with decreasing activity \citep{Petit2008}, the $B_l$ may be unable to reliably recover the field due to cancellations between small-scale, oppositely polarized regions. These scenarios could explain why we are unable to recover any increase in $B_l$ as $\log R'_{\rm HK}$ increases between -4.8 and -4.4. However, they do not explain why we are able to recover increasing magnetic field strength with increasing chromospheric activity for stars with $\log R'_{\rm HK}\leq-4.8$. For all inactive stars, we would instead expect to see a vertical stacking of stars having continually decreasing chromospheric activity and an almost constant, minimum detectable $B_l$. 

Perhaps the strongest argument against a systematic cause for the three regimes in the $\log R'_{\rm HK}$ - $\log |B_l|$ relation, is the fact that we also see a similar and possibly related `step' in the relationship between chromospheric activity variability amplitudes and mean chromospheric activity, which we discuss further in section \ref{sec:var-sg,ms,pms}.  If the distinct regimes in the $\log R'_{\rm HK}$ - $\log |B_l|$ relation are real, they may relate to a change in the surface properties of the magnetic dynamo at around $\log|B_l|\sim0.4$. One possibility is a change in the area ratio of chromospheric plages to photospheric spots, as suggested by \citet{Foukal2018}. \citet{Foukal2018} proposed that stellar plage/spot ratios may be lower for stars that are slightly more active compared to the Sun ($\log R'_{\rm HK}\sim-4.8$ in cycle 23, \citet{Lehmann2021}), based on the fact that the Sun exhibits a reduced plage/spot area ratio when at its most active \citep{Foukal1998}. This phenomenon could also relate to the well-observed change from spot-dominated to faculae-dominated photospheric variability with decreasing mean chromospheric activity and increasing stellar age \citep{Radick1998,Lockwood2007,Radick2018}, which occurs at $\log R'_{\rm HK}\sim -4.65$ to -4.75. If stars with $\log|B_l|\sim0.4$ are impacted by a reduced plage/spot ratio then they would appear weaker in terms of their chromospheric activity (which relates to chromospheric plages) relative to their $|B_l|$ (which relates to the magnetic fields in both spots and faculae). 
The affected stars would drop vertically down in our Figure \ref{fig:SvBl}, which would be consistent with our observations.  Interestingly, \citet{Nichols-Fleming2020} observed a similar, sudden decrease in X-ray luminosity variability (which is known to correlate strongly with chromospheric \ion{Ca}{ii} H\&K emissions) for G stars, which occurred at a rotation period of $\sim$15 d, and this rotation period also coincides with our observed $\log R'_{\rm HK}$ `step down' for G stars (see Figure \ref{fig:AppB}). 

\subsubsection{Trends with stellar age and spectral type}\label{sec:trends with stellar age and spectral type}

For both G and K stars, chromospheric activity and magnetic field strengths are higher for young stars (open circles) compared to the more mature main sequence stars (filled circles). For F and M-type stars, there appears to be a similar decrease in activity and magnetic field strengths from young to mature stars, but our sample size, particularly the samples of young stars, are small compared to G and K stars. As a result, the distributions of chromospheric activity and magnetic field strengths for young F and M stars are dominated by statistical noise. 

The lower-limits of chromospheric activity and magnetic field strengths for young G and K stars are similar ($\log R'_{\rm HK} \sim-4.3$ and $\log |B_l|\sim1.0$), but the young K stars extend to higher mean $\log R'_{\rm HK}$ and $\log|B_l|$ compared to the G stars. This is likely to be related (at least partly) to a sample bias; young and highly active K (and M) stars are targeted in priority for spectropolarimetric observations to increase the likelihood of a magnetic detection.  The population of young stars appears to correspond to the `very-active' group of stars with $\log R'_{\rm HK}\geq-4.2$ observed by \citet{Gomes2020}. 

For both G and K stars, the populations of young stars are separated in both the $\log R'_{\rm HK}$ and $\log|B_l|$ histograms from the populations of mature, main sequence stars, but again it is possible that the apparent grouping of stars is an effect of targeted observing programs with NARVAL and ESPaDOnS. If it is a real phenomenon, the under-density of stars between the young and main sequence populations resembles the `gap' in stellar rotation periods described by \citet{Barnes2003} that occurs around the transition from the pre-main sequence to the main sequence. \citet{Barnes2003} propose that the core and envelope of a star re-couple in this region, and the magnetic field may change from a `convective' field to an `interface' field produced at the tachocline. The presence of the gap in both the chromospheric activity and magnetic field strengths is consistent with a rapid evolution between these pre-main sequence and main sequence phases. 

For main sequence F, G and K stars, the lower and upper-limits of chromospheric activity and magnetic field strengths are similar, while our sample of M stars appears to generally have higher $\log R'_{\rm HK}$ and $\log|B_l|$. Again, this may be partly related to a sample bias, whereby active M stars are preferentially targeted. Or it could be related to the longer main sequence lifetimes of mid-M stars, their less-efficient activity/rotation spin down or the difference in convective turnover timescales between stellar types. 

The $\log R^{\prime}_{\rm HK}$ histograms for main sequence F, G and K stars (grey shaded histograms) show that chromospheric activity may be bimodally distributed, or could be skewed toward inactive stars with an extended tail toward higher activity stars. Figure \ref{fig:SvBl} shows slightly under-populated regions of F stars between $\log R'_{\rm HK}=-4.7$ and -4.8, G stars between -4.4 and -4.8, and K stars between -4.5 and -4.7. In each case, the under-density of stars is not significant for our sample when we consider the uncertainties in the histogram bins (Figure \ref{fig:AppA}), but a bimodal distribution is significant for G stars when we combine our data with that from \citet{BoroSaikia2018} and \citet{Gomes2020}. Similar to \citet{BoroSaikia2018}, the possible reduced population of intermediately--active G stars in Figure \ref{fig:AppA} is less prominent compared to the `Vaughan-Preston gap', originally shown by \citet{Noyes1984} to be completely devoid of F and G stars between $\log R^{\prime}_{\rm HK}=$-4.6 and -4.9.  This is likely due to our larger sample size compared to \citet{Noyes1984}. 
We do not observe any distinct, `very-inactive' group of stars in any of the chromospheric activity distributions, with $\log R'_{\rm HK}\leq -5.0$, as was detected by \citet{Gomes2020}.  This is probably because their data includes stars evolving off the main sequence, while our sample excludes these stars.

In contrast to the chromospheric activity distributions, the $\log|B_l|$ data for main sequence stars seem to be normally distributed, particularly across each of the G and K samples. It is interesting that the low-population region that separates young stars from old for both the G and K samples, is evident in both $\log|B_l|$ and $\log R'_{\rm HK}$ distributions, while the reduced population of main sequence G stars with intermediate activity is shown in only the $\log R'_{\rm HK}$ distribution. The fact that we see no discontinuity in $\log|B_l|$ for G stars at intermediate activity supports a change in dynamo properties on the main sequence, such as those we discussed in section \ref{sec:gradient_change}, rather than a rapid evolutionary phase.

\subsection{Chromospheric activity and magnetic field strength variability amplitudes}\label{sec:var_amplitudes}

\begin{figure}
    \centering
    \includegraphics[width=0.95\linewidth]{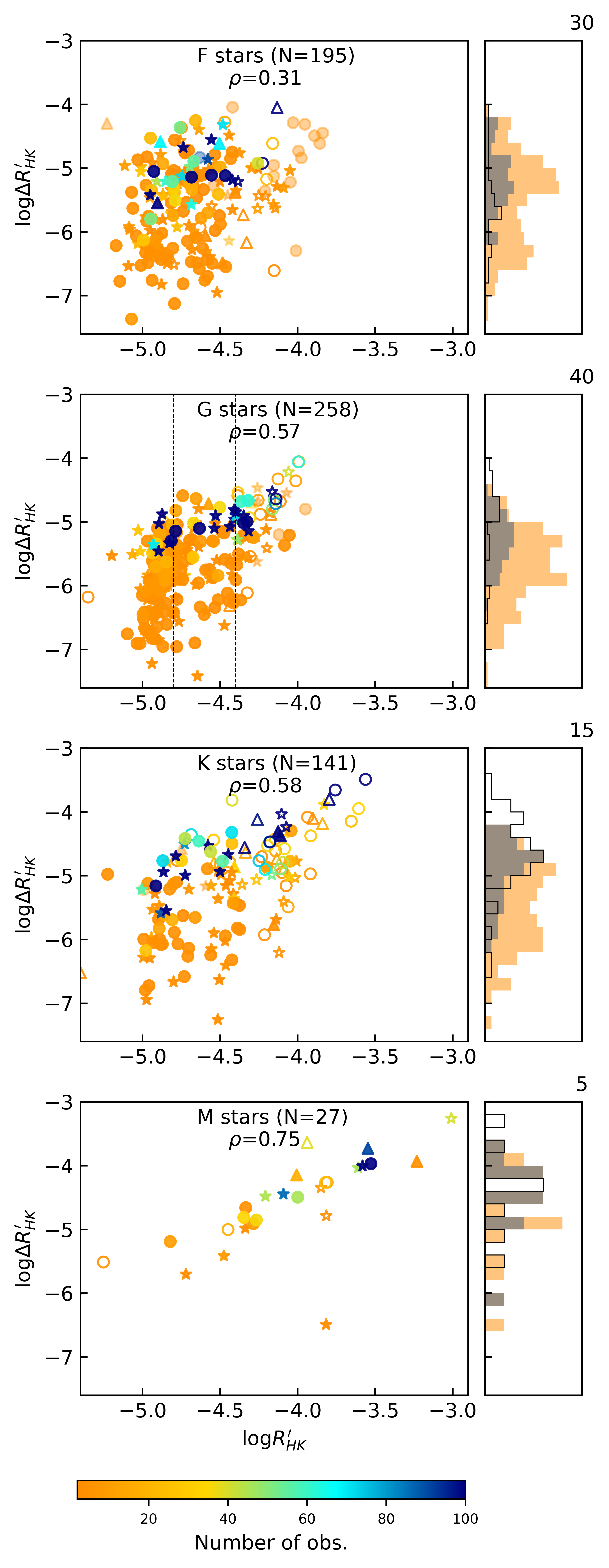}
    \caption{Amplitude of chromospheric activity variability versus mean chromopsheric activity for samples of F, G, K and M stars. Marker shapes and fill-styles are the same as in Figure \ref{fig:HRD}. Marker color scales with the number of observations. The histograms to the right of the plots indicate the distributions of $\log \Delta R^{\prime}_{\rm HK}$ for the entire main sequence sample (orange), main sequence stars with $\geq10$ observations (grey), and young stars (black line). The dashed lines correspond to the reduced population of G stars from Figure \ref{fig:SvBl}. }
    \label{fig:SvBl2}
\end{figure}

\begin{figure*}
    \centering
    \includegraphics[width=0.8\linewidth]{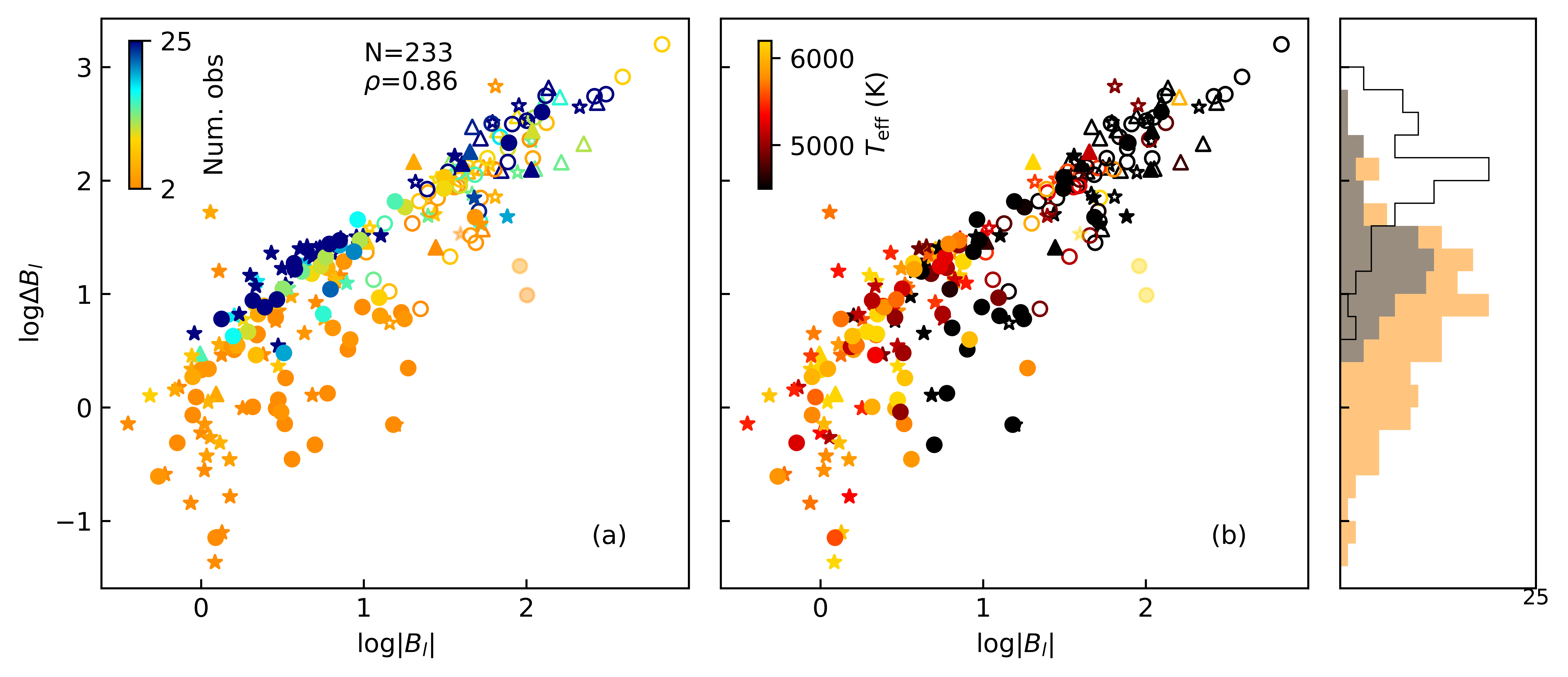}
    \caption{Magnetic field variability amplitude versus mean magnetic field strength. Marker shapes and fill-styles are the same as in Figure \ref{fig:HRD}. In (a) the marker color scales with the number of observations, while in (b) marker color scales with stellar effective temperature. The histograms to the right of the plots indicate the distributions of $\Delta B_l$ for the entire main sequence sample (orange), main sequence stars for which the variability amplitude is based on $\geq10$ observations (grey), and young stars (black line). }
    \label{fig:BvarVBl}
\end{figure*}

Figures \ref{fig:SvBl2} and \ref{fig:BvarVBl} compare the peak-to-peak amplitudes of $\log R^{\prime}_{\rm HK}$ and $\log B_l$ variability with the mean $\log R^{\prime}_{\rm HK}$ and mean $\log |B_l|$ respectively. In Figure \ref{fig:SvBl2}, stars are grouped by spectral type, and the marker color scales with the number of observations. In Figure \ref{fig:BvarVBl} we show all spectral types together, and in \ref{fig:BvarVBl}(a) marker color scales with the number of observations, while in (b) it indicates stellar temperature. As in Figure \ref{fig:SvBl}, the number of stars in each sub-sample and the Pearson correlation coefficient are shown at the top of each plot.  Filled markers represent mature stars, open markers show young stars and the black line histograms indicate the distributions of activity and magnetic field strength for the young sample. Due to the significant scatter in these plots, we show the distributions of $\log \Delta R'_{\rm HK}$ and $\log \Delta B_l$ for each entire main sequence sample in orange. The grey histograms relate to filtered samples of main sequence stars that have $\geq10$ observations. 

\subsubsection{$\log \Delta R'_{\rm HK}$ versus mean $\log R'_{\rm HK}$}\label{sec:var-sg,ms,pms}

For G, K and M stars, the amplitude of $\log R^{\prime}_{\rm HK}$ variability shows a moderate, positive correlation with the mean $\log R^{\prime}_{\rm HK}$. \citet{Saar2002} similarly observed an increase in chromospheric activity amplitudes with mean activity above $\log R'_{\rm HK}=-5.0$. Meanwhile, for F stars there is only a weak correlation between chromospheric activity variability and the mean activity level. The positive correlations for G and K stars are strongest for stars with a high number of observations. Unlike in Figure \ref{fig:SvBl}, where stars with few observations are scattered both above and below the main trend, it is clear in Figure \ref{fig:SvBl2} (and \ref{fig:BvarVBl}) that stars with few observations are scattered mostly below the trend.  This is because variability amplitudes are inevitably underestimated for poorly observed stars. Some of the scatter will also be due to the fact that activity is restricted in latitude, such that stars with different inclinations to the observer will have different apparent activity amplitudes \citep{Saar2002}. 

Considering targets with at least $\sim 50$ obs. (green to dark blue markers), there appear to be at least two regimes of activity variability for G stars, and possibly for K stars. G stars show decreasing chromospheric activity variability with decreasing mean activity between $\log R'_{\rm HK}\sim -3.8$ and $-4.4$, but below $\log R'_{\rm HK}\sim -4.4$ there is only a minimal change in chromospheric activity variability with decreasing mean activity. It is possible that chromospheric variability again begins to decrease with mean activity below $\log R'_{\rm HK}\sim-4.8$, but this is based on a very small sample of well-observed stars. Our chromospheric activity variability data combined with the results from \citet[][Figure \ref{fig:AppC}]{Gomes2020} are also consistent with a non-linear relationship between $\log \Delta R^{\prime}_{\rm HK}$ and $\log R'_{\rm HK}$. Although the functional form of the relationship is poorly constrained, the combined data set supports multiple regimes of chromospheric activity variability. \citet{Gomes2020} interpreted their results as three possible chromospheric variability regimes but they found that only the upper envelope of chromospheric activity variability scales with mean activity across all regimes. Given that the significant scatter in Figure \ref{fig:SvBl2} seems to be mostly related to stars with low numbers of observations, we find that both the upper and lower levels of activity variability scale well with mean activity. 

It is interesting that the possible changes in the $\log \Delta R'_{\rm HK}$ - mean $\log R'_{\rm HK}$ relation seem to occur at a similar activity level as changes in the mean $\log R'_{\rm HK}$ - $\log|B_l|$ relation, and may coincide with the reduced-population region of intermediately-active G stars from Figure \ref{fig:SvBl} (indicated by dashed lines in Figure \ref{fig:SvBl2}). This suggests that the change in the mean $\log R'_{\rm HK}$ - $\log|B_l|$ relation is not related to some systematic over or underestimation of $|B_l|$ for low-activity stars. Rather, it supports a change in the properties of the magnetic dynamo at the stellar surface.

For our samples of G and K--type stars, the mature main sequence stars generally have lower mean chromospheric activity and variability compared to the more youthful stars. Variability amplitudes appear to have similar ranges across spectral types F to K, for both the young and main sequence samples, and the variability amplitudes of young stars are not distinctly separated from the variability of mature stars. The chromospheric activity variability of main sequence F, G and K stars appears to be normally distributed, similar to the $\log|B_l|$ distributions shown in Figure \ref{fig:SvBl}. There is no reduced population in the $\log \Delta R'_{\rm HK}$ distribution that corresponds to the reduced population of main sequence G stars with intermediate mean chromospheric activity in Figure \ref{fig:SvBl}.  Although the orange histograms in Figure \ref{fig:SvBl2} show possibly bimodal distributions of $\log \Delta R'_{\rm HK}$, this is clearly related to the scattering of data below the main trend. 

\subsubsection{Magnetic field variability versus mean field strength}

Figure \ref{fig:BvarVBl} shows a strong, positive correlation between $\log \Delta B_l$ and mean $\log |B_l|$ for well observed stars (green to dark blue markers). Young stars typically have stronger magnetic fields with greater variability amplitudes compared to mature stars, and there is clear separation between our samples of young and mature stars in the $\log \Delta B_l$ distribution. K and M stars generally show higher magnetic field strengths and magnetic field variability compared to the warmer F and G stars, but this is somewhat impacted by the fact  that active K and M stars are preferentially targeted for spectropolarimetric observations. 
 
\subsection{Chromospheric activity versus magnetic field strength from ZDI}

\begin{figure}
    \centering
    \includegraphics[width=0.9\columnwidth]{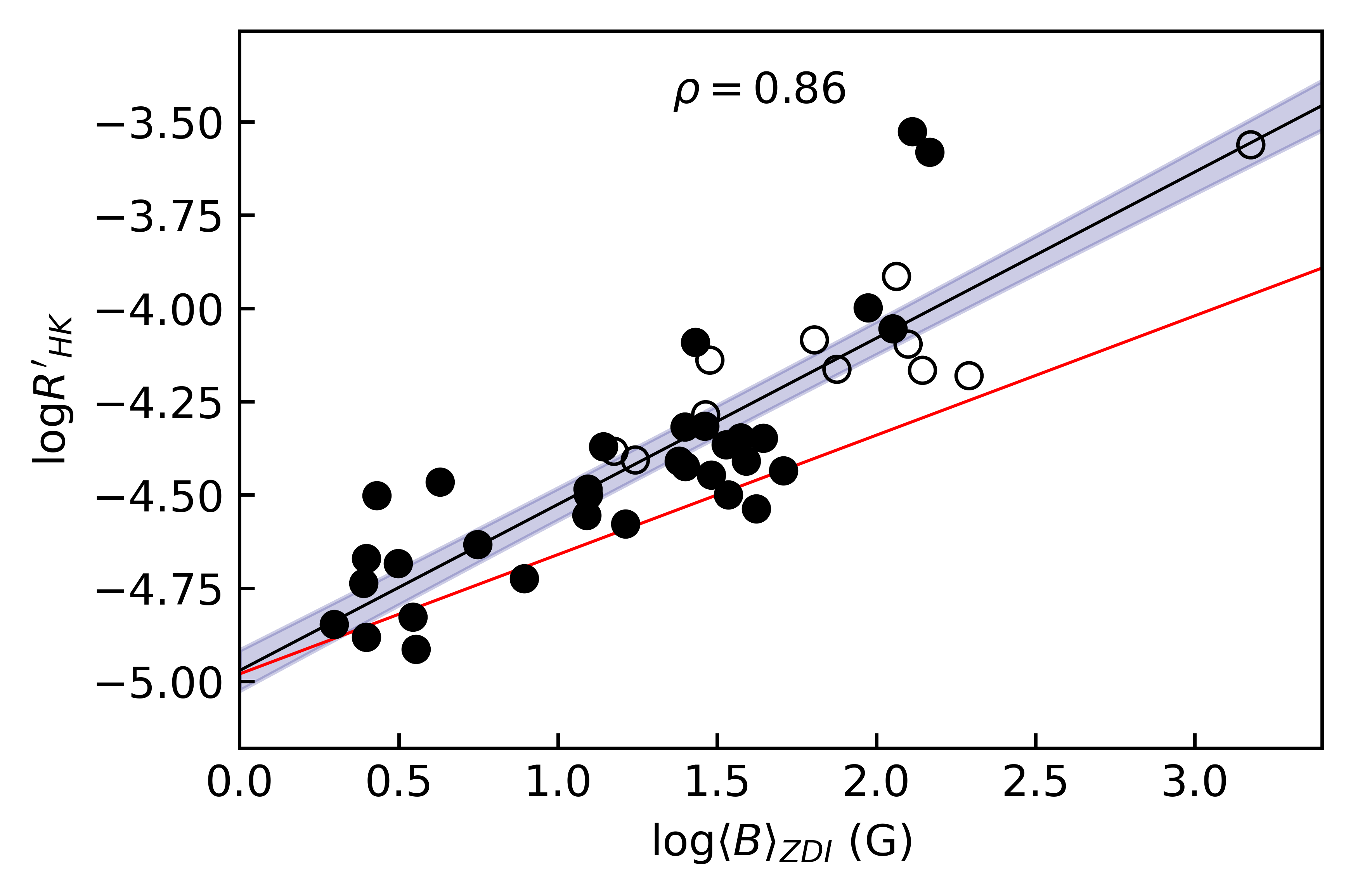}
    \caption{Mean chromospheric activity from our study versus surface-averaged large-scale magnetic field strength from published ZDI work. Marker fill--styles are the same as in Figure \ref{fig:HRD}; open circles represent young stars with strong spectral \ion{Li}{i} lines and filled circles are mature stars assumed to be on the main sequence. The black regression line has a slope of 0.44 $\pm$ 0.04, and the grey shaded region shows its 95\% confidence interval. The red line shows the previous results of \citet{Petit2008}.}
    \label{fig:ZDI5}
\end{figure}

Figure \ref{fig:ZDI5} compares the surface-averaged large scale magnetic field strength from published ZDI maps\footnote{ 
\citet{BoroSaikia2015,BoroSaikia2016,BoroSaikia2018a,Brown2021,doNascimento2016,Donati2008,Donati_BPtau,Donati2008a,Donati_AATau,Donati2129Oph,Donati_TWHya,Donati_DNTau,Donati_LKCA,Donati2015,Fares2009,Fares2010,Fares2012,Fares2013,Folsom2016,Folsom2018a,Folsom2018,Folsom2020,Hebrard2016,Jeffers2014,Jeffers2017,Jeffers2018,Marsden2006,Mengel2016,Morgenthaler2011,Morin2008a,Morin2010,Petit2008,See2019,Waite2015,Waite2017}}
to the mean $\log R^{\prime}_{\rm HK}$ activity we calculated from PolarBase observations. For stars that have been observed using ZDI over multiple epochs, we show the mean field strength from all ZDI maps. We have removed outliers by excluding stars for which the mean $\log R^{\prime}_{\rm HK}$ is derived from $\leq 20$ Stokes {\it{I}} observations.  

The data indicate that mean chromospheric activity is directly related to the strength of the large-scale magnetic field recovered from ZDI. This is consistent with previous work by \citet{Petit2008}, although they found a lower slope, $R^{\prime}_{\rm HK} \propto \langle |B| \rangle_{\rm ZDI} ^{0.33}$, compared to $R^{\prime}_{\rm HK} \propto \langle |B| \rangle _{\rm ZDI}^{0.44 \pm 0.04}$ for our data. If we consider only main sequence F-K stars we obtain a slope of $\sim0.29$, closer to the results of \citet{Petit2008}. 
The relationship between $\log R'_{\rm HK}$ and $\log \langle |B| \rangle _{\rm ZDI}$ is consistent with the dependence of $\log R'_{\rm HK}$ on the surface-averaged longitudinal magnetic field shown in Figure \ref{fig:SvBl}, although we do not find evidence of a change in slope between $\log R^{\prime}_{\rm HK}$ and $\log \langle |B| \rangle _{\rm ZDI}$. This is probably due to the small sample size in Figure \ref{fig:ZDI5} compared to Figure \ref{fig:SvBl}. 

\subsection{Chromospheric activity and magnetic field geometry versus effective temperature}\label{sec:rhk_v_teff}

\begin{figure*}
    \centering
    \includegraphics[width=0.8\linewidth]{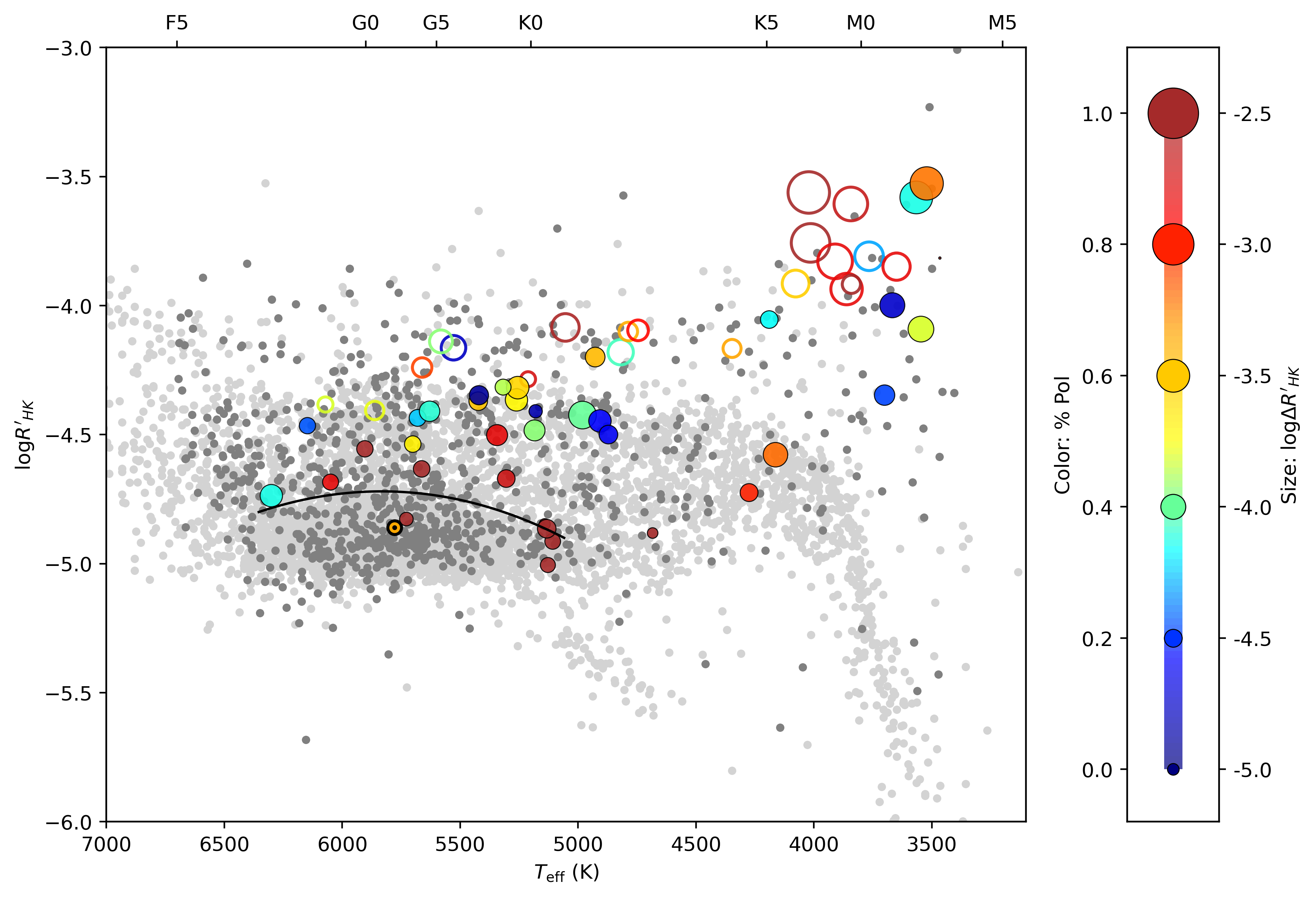}
    \caption{Mean chromospheric activity versus stellar effective temperature for our sample of stars (dark grey) and published data (light grey) from \citet{BoroSaikia2018} and \citet{Gomes2020}. The properties of the large-scale magnetic field are shown for a selection of stars with published ZDI maps. Marker color represents the fraction of the large-scale magnetic field that is stored in the poloidal component and marker size represents our measured activity variability amplitude, $\log \Delta R^{\prime}_{\rm HK}$. Marker fill-styles are the same as in Figure \ref{fig:HRD}, and the Sun is indicated with the usual symbol. The black curve corresponds to the black curve indicated in Figure \ref{fig:ZDI2}.}
    \label{fig:ZDI1}
\end{figure*}

\begin{figure*}
    \centering
    \includegraphics[width=0.7\linewidth]{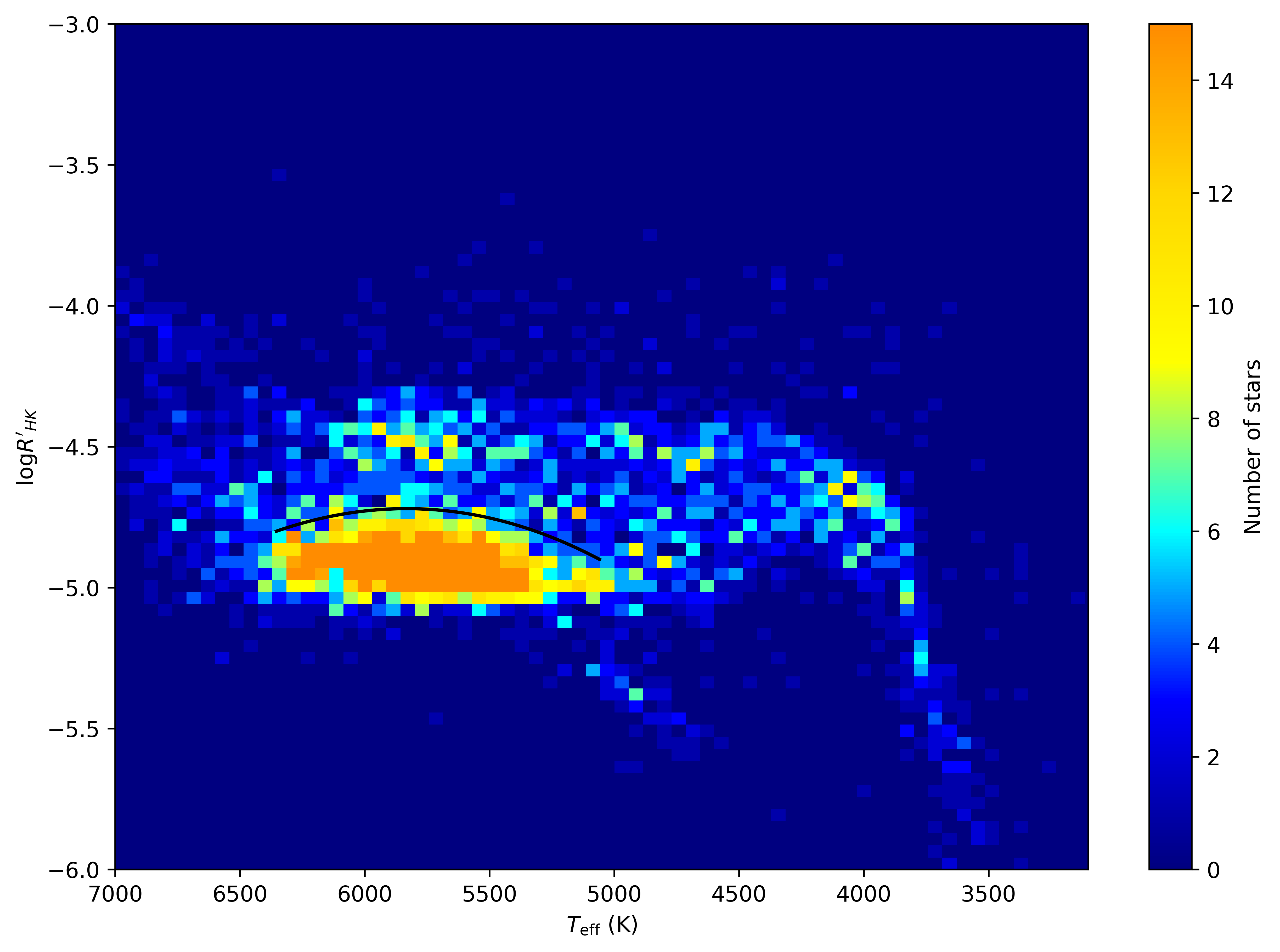}
    \caption{Two-dimensional histogram indicating the number of stars in Figure \ref{fig:ZDI1} populating equal--sized temperature and $\log R^{\prime}_{\rm HK}$ bins. For targets that are common to both PolarBase and the published data we take the mean of all available $\log R^{\prime}_{\rm HK}$ values for Figure \ref{fig:ZDI2}. The black curve we have drawn is intended to draw the eye to the densely population region we discuss in section \ref{sec:rhk_v_teff}, and corresponds to the black curve shown in Figure \ref{fig:ZDI1}.}
    \label{fig:ZDI2}
\end{figure*}

Figure \ref{fig:ZDI1} shows mean chromospheric activity versus stellar effective temperature for our sample (dark grey markers). We have also included the publicly available chromospheric activity data from \citet{Gomes2020}, and activity data compiled by \citet{BoroSaikia2018} (light grey markers). Activity amplitudes (symbol size) and the poloidal fraction of the large-scale magnetic field (symbol colour) are also shown in Figure \ref{fig:ZDI1} for a selection of stars with published ZDI maps.  We show the Sun with the usual symbol, which is scaled for its activity variability during cycle 23 \citep{Lehmann2021}. The fractional poloidal field of the Sun we show here is the minimum value measured by \citet{Lehmann2021}; note that since the Sun's magnetic field has been observed in much higher cadence compared to the other stars, its large-scale toroidal component has been better recovered. For the combined sample shown in Figure \ref{fig:ZDI1}, we also computed a two-dimensional histogram which is shown in Figure \ref{fig:ZDI2}. 

\subsubsection{Young stars}

Youthful stars, shown by open circles, are clearly more active compared to main sequence stars across all temperatures, and have greater activity variability, as has been shown in the previous sections. In the upper right of  Figure \ref{fig:ZDI1} there are a number of very-cool, highly active stars that are marked as main sequence age (filled circles), for which we did not detect strong \ion{Li}{i} lines within the spectra. According to {\sc{simbad}} these are all known to be young, pre-main sequence stars (OT Ser, DS Leo, BD+132618, BD+61195 and HD209290). 

Young stars with $\log R^{\prime}_{\rm HK}\geq -4.0$ and $3800\leq T_{\mathrm{eff}}\leq4000$\,K appear to have strongly poloidal magnetic field geometries. This contrasts against young, active stars with $T_{\mathrm{eff}}\leq3800$\,K, which show mixed field geometries and comparatively lower variability in chromospheric activity. \citet{Morin2011} suggested that the presence of both dominantly toroidal and dominantly poloidal fields in active M stars could indicate dynamo bi-stability. Another interpretation is that such stars have similar magnetic cycles to the solar cycle, and that those observed to have dominant toroidal fields have simply been observed at times of a magnetic field inversion, which is when the solar magnetic field becomes its least poloidal \citep{Kitchatinov2014}.

Young stars with $T_{\mathrm{eff}}\geq4000K$ also show mixed field geometries, and have $\log R^{\prime}_{\rm HK}$ between $\sim-4.0$ and -4.3. The change in field geometry at $\sim 4000K$ from poloidal to mixed geometries has been well observed \citep{donati2011c,Gregory2012,Folsom2018a,Hill2019}, and is thought to occur around the transition from fully-convective to partially-convective stars \citep{Folsom2018a,Hill2019,Villebrun2019}. For the young stars with $T_{\mathrm{eff}}\geq4000K$ there is no clear dependence of the strength of the poloidal field on stellar temperature or mean activity level. This is also shown in Figure \ref{fig:ZDI4}, where we directly compare $\log R^{\prime}_{\rm HK}$ and $\log \Delta R^{\prime}_{\rm HK}$ to the fraction of the magnetic field stored in the poloidal component, and the fraction of the poloidal field stored in high order modes. For young stars, the poloidal field may reach a possible minimum fraction of 20-40 percent at around $\log R^{\prime}_{\rm HK}\sim -4.25$, before stars reach the main sequence.

\subsubsection{Main sequence stars}

For main sequence stars with $4200\leq T_\mathrm{eff} \leq 6300$\,K, the upper level of chromospheric activity has little dependence on $T_\mathrm{eff}$. This suggests that late-F to late-K stars begin their main sequence lives with a similar level of chromospheric activity, $\log R^{\prime}_{\rm HK}\sim-4.3$. The basal level of chromospheric activity also appears to be independent of $T_\mathrm{eff}$ for $4500\leq T_\mathrm{eff} \leq 6500$\,K, apart from a small very-low activity group of stars at around 5000\,K, which are from the \citet{Gomes2020} sample and are likely to be subgiant stars evolving off the main sequence. The basal activity increases for stars with $T_\mathrm{eff}\approx4500$ toward 4200 K. \citet{Mittag2013} found a similar lack of low activity stars in the B-V range of 1.1 to 1.5. A simple explanation for this may be that K and M stars have different convective properties compared to F and G stars, and/or longer main sequence lifetimes with less--efficient loss of activity as they spin-down, such that they have not reached a similarly low activity level.  For $T_{\mathrm{eff}}\leq 4200$\,K, the upper and lower activity levels for main sequence stars decrease toward lower temperatures. It is not clear why these very cool stars begin the main sequence with lower chromospheric activity compared to F, G and K stars, but decreasing $\log R'_{\rm HK}$ with decreasing $T_\mathrm{eff}$ is consistent with previous work \citep{Mohanty2003,Reiners2008,Astudillo2017}.   For main sequence stars with $T_\mathrm{eff}\geq 6300$\,K, the population of very low activity stars  ($\log R^{\prime}_{HK}\sim -5.0$) begins to disappear. It is possible that this is because the chromospheric activity for such low-activity stars is difficult to detect, or could be related to differences in convective properties for F stars. 
It should also be noted that the stars with $T_\mathrm{eff}\geq 6300$\,K and $\leq 4200$\,K are mostly taken from \citet{BoroSaikia2018}, and extend beyond the S-index to $\log R^{\prime}_{\rm HK}$ calibration range we use here, which should be taken into account when comparing trends. 

\begin{figure*}
    \centering
    \includegraphics[width=0.7\linewidth]{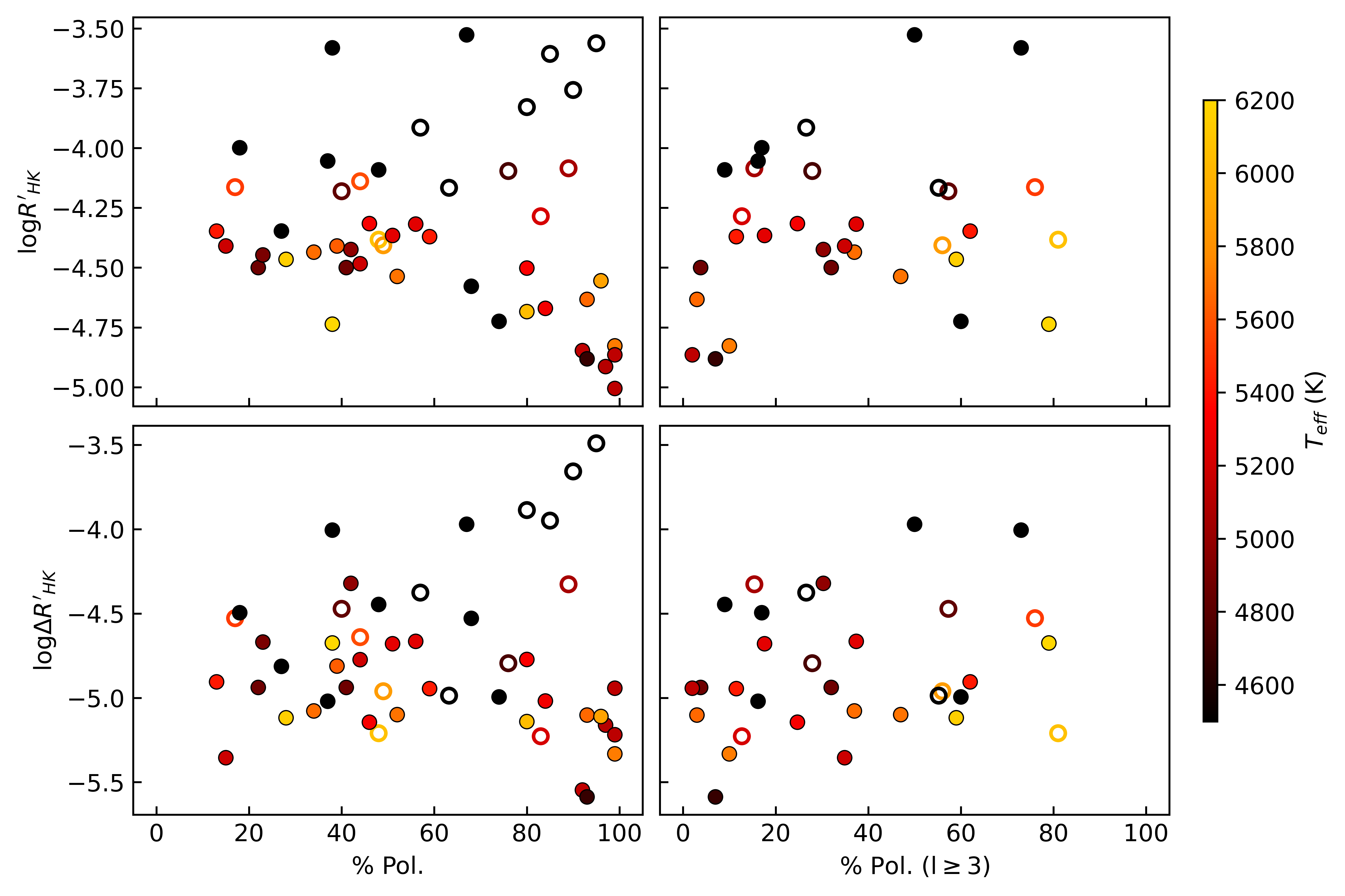}
    \caption{Left: $\log R^{\prime}_{\rm HK}$ and $\log \Delta R^{\prime}_{\rm HK}$ versus the fraction of the large scale magnetic field stored in the poloidal component. Right: $\log R^{\prime}_{\rm HK}$ and $\log \Delta R^{\prime}_{\rm HK}$ versus  the fraction of the poloidal field stored in octopolar or higher energy modes. Marker color scales with stellar effective temperature, and marker styles are the same as in Figure \ref{fig:HRD}. We have removed outliers by excluding stars for which $\log R^{\prime}_{\rm HK}$ and $\log \Delta R^{\prime}_{\rm HK}$ are derived from $\leq 20$ high SNR observations. }
    \label{fig:ZDI4}
\end{figure*}

The ZDI results show clear evolution of the large-scale surface magnetic field on the main sequence. Figures \ref{fig:ZDI1} and \ref{fig:ZDI4} indicate that for main sequence F, G and K stars, the poloidal component of the magnetic field increases as the mean $\log R^{\prime}_{\rm HK}$, and possibly the $\log \Delta R^{\prime}_{\rm HK}$, decrease. This is consistent with previous studies, which showed that both the toroidal energy fraction \citep{Donati2009,Folsom2016,See2016} and chromospheric activity cycle amplitudes \citep{Saar2002} decrease with the Rossby Number ($Ro$, ratio of rotation period to convective turnover time, see Appendix \ref{fig:AppD}). Figure \ref{fig:ZDI4} also shows a wide range in the complexity of the poloidal field component for main sequence F-K stars, from fairly simple fields with almost all of the poloidal field energy stored in octopolar or lower modes, to as much as 80 percent of the poloidal field stored in higher order modes.

The late-F to late-G stars shown in Figure \ref{fig:ZDI1} appear to transition at $\log R^{\prime}_{\rm HK}\sim -4.5$ from being capable of generating significant toroidal fields to having dominantly poloidal fields. Our data are also consistent with K stars undergoing a shift from significantly toroidal to dominantly poloidal fields at a similar activity level to G and late-F stars, although published ZDI maps are sparse for K stars around $\log R^{\prime}_{\rm HK}=-4.5$. For mid-F stars, strong toroidal fields may prevail toward lower activity levels, suggesting that the transition to dominantly poloidal fields may occur at a lower $\log R^{\prime}_{\rm HK}$. It is not clear from our sample if M stars transition to dominantly poloidal field structures as their chromospheric activity and activity variability decrease. 

The 2-dimensional histogram in Figure \ref{fig:ZDI2} indicates a densely populated, low activity region with $5000\leq T_{\mathrm{eff}} \leq 6300$\,K and $-5.05 \leq \log R^{\prime}_{\rm HK} \leq -4.75$ (marked by a black curve in Figures \ref{fig:ZDI1} and \ref{fig:ZDI2}). Above this region, there is a slightly reduced population of late-F to early-K stars with activity between -4.50 and -4.75. The reduced population is less-prominent compared to the original Vaughan-Preston gap, which is consistent with the findings of  \citet{BoroSaikia2018}. Above the reduced population, there is also a slightly higher-density band of active main sequence stars, with $\log R'_{\rm HK}$ between $\sim -4.30$ and -4.50. This band covers a smaller temperature range compared to the densely populated region of inactive stars (marked by the black curve), and may extend from $\sim6100$ to 5300\,K.  These features are consistent with the bimodal distribution of main sequence G stars shown in Figure \ref{fig:AppA}, and suggest that the bimodal distribution could extend to slightly higher and lower temperatures to include late-F to early-K stars. For stars with $T_{\rm eff}$ below $\sim5000$\,K, the distribution of activity appears to be fairly flat and dominated by statistical noise. \citet{Gomes2020} observed a triple-peaked distribution in their sample of stars with $4350\leq T_{\rm eff}\leq5280$\,K (which they classify as K stars), but we do not find this to be clearly evident in Figure \ref{fig:ZDI2}, nor significant with respect to the histogram uncertainties in Figure \ref{fig:AppA}.  

Similar to the findings of \citet{Metcalfe2016}, our results indicate that the transition from strong toroidal fields to dominantly poloidal fields occurs at a similar activity level as the upper-boundary of the under-populated region of intermediately-active main sequence stars we identified in Figure \ref{fig:ZDI2}. \citet{Metcalfe2016} suggested that the underpopulated region could be explained by a shift in the character of differential rotation, possibly resulting in a period of rapid evolution. Considering that the under-density also occurs at a similar activity level as the possible changes in the relationships between mean activity, magnetic field strength and activity variability that we described in the previous sections, we find that the under-populated region is more likely related to a change in magnetic surface properties, such as a change in the plage-spot ratio, rather than a period of rapid evolution.  

\subsection{Chromospheric activity and magnetic field geometry versus rotation period}

\begin{figure}
    \centering
    \includegraphics[width=0.85\columnwidth]{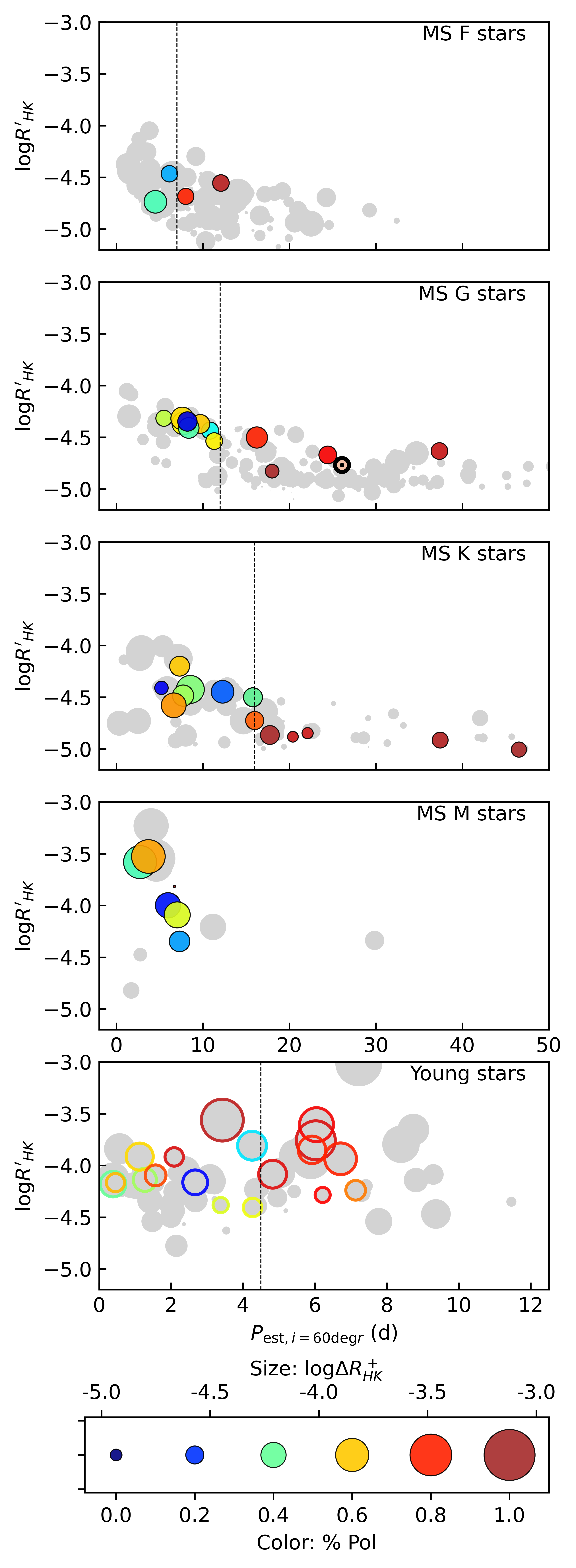}
    \caption{Mean $\log R^{\prime}_{\rm HK}$ versus the estimated stellar rotation period for our sample of stars (grey). Rotation periods were derived from the published stellar radius and $v\sin{i}$ from Table \ref{tab:stellar_properties} and assuming an inclination angle of $60\degr$. The properties of the large-scale magnetic field are also shown for a selection of stars with published ZDI maps. Markers are the same as in Figure \ref{fig:ZDI1}. Note that for young stars the plot shows a reduced range of rotation periods. The vertical dashed lines indicate the estimated transition from stars capable of generating significant toroidal field geometries, to those with dominantly poloidal fields. }
    \label{fig:ZDI3}
\end{figure}

Figure \ref{fig:ZDI3} compares chromospheric activity, activity variability (symbol size) and magnetic field geometry (symbol colour) to the stellar rotation period for subsets of main sequence F, G, K and M stars, as well as a separate plot for young, \ion{Li}{i} abundant stars. Note that the rotation periods used here are estimates only, based on our derived stellar radius, $v\sin{i}$ and assuming an inclination angle of 60$\degr$. 

The data show that mean activity decreases as stellar rotation slows for all stars on the main sequence. This is consistent with stellar spin-down theory \citep{Skumanich1972}, where young stars have high magnetic field strengths and chromospheric activity, which diminish throughout their main sequence lifetimes due to angular momentum loss from the coupling of the magnetic field and stellar wind. The chromospheric activity of F, G and K stars appears to plateau at $\log R^{\prime}_{\rm HK}\approx-5.0$ for slower rotating stars. This levelling-off of chromospheric activity as rotation slows is not obvious for our sample of M stars, possibly because our sample is biased toward active M stars, or because low mass M stars have longer spin--down timescales \citep{West2008,Johnstone2021}. \citet{Petit2008} found for their sample of G stars that a rotation period lower than $\sim$12\,d is necessary for the toroidal magnetic energy to dominate over the poloidal component. This is consistent with our results for main sequence G-type stars. Our data also suggest that dominant toroidal fields are only present for main sequence F stars with estimated rotation periods below $\sim7$d, and for main sequence K stars with estimated rotation periods below $\sim16$d. This is likely to be related to convection zone depth, which decreases from K to F stars. The change from strongly toroidal to dominantly poloidal fields is hypothesized to occur when the rotation period becomes close to the convective turnover time ($Ro\sim1$, see Appendix \ref{fig:AppD}), which would occur at shorter rotation periods for F stars with shallow convective zones, and at longer rotation periods for later-type stars with deeper convection zones. 

Young stars with greater chromospheric activity and variability, and dominantly poloidal field structures, have longer rotation periods compared to stars those with weaker activity variability and which are capable of generating both dominantly poloidal and toroidal fields. This is consistent with previous work on young stars \citep{Folsom2018a,Hill2019,Villebrun2019}, which indicate that magnetic fields become more complex as young stars contract toward the main sequence and their rotation rate increases. The transition from dominantly poloidal to significantly toroidal field structures appears to occur at an estimated rotation period of $\sim4.5$ d.   

\section{Conclusions}

We provide a catalog of new chromospheric activity and surface-averaged large-scale magnetic field measurements for 954 main sequence and youthful stars with spectral types from mid-F to mid-M. This includes mean values and peak-to-peak variability amplitudes for both the $\log R^{\prime}_{\rm HK}$ and $B_l$, which were derived using time-series spectropolarimetric observations from the PolarBase database. In this work we compare chromospheric activity, magnetic field strengths and their variability amplitudes. We also use previously published chromospheric activity data and ZDI results to diagnose relationships between chromospheric activity, activity variability, large-scale magnetic field geometries and stellar properties. 

The data confirm that chromospheric activity, magnetic field strengths, and their variability amplitudes are generally higher for youthful stars compared to more mature, main sequence stars. Stars from late-F through to mid-K type appear to have similar levels of activity and similar magnetic field strengths as they begin their main sequence lives. For F, K and M stars, chromospheric activity decreases fairly smoothly with decreasing mean, unsigned magnetic field strength. Conversely for G stars, our data suggest that the $\log R'_{\rm HK}$ - $\log|B_l|$ relationship may have three distinct phases; mean chromospheric activity and field strength decrease together until $\log R'_{\rm HK}\sim-4.4$ and $\log|B_l|\sim0.4$, at which point there is a significant step down in chromospheric activity but minimal change in magnetic field strength. This is followed by another decreasing phase for both chromospheric activity and magnetic field strength for $\log R'_{\rm HK}\leq-4.8$ and $\log|B_l|\leq0.4$. 
The distinct phases in the $\log R'_{\rm HK}$ and $\log|B_l|$ relationship could be related to a change in the surface properties of the magnetic dynamo on the main sequence, such as a change in the area ratio of plages and spots. Further long-term observations of the magnetic fields and chromospheric activity of stars across a range of activity levels will be required to determine if the three phases we observed are real, and if they are unique to G stars. 

The amplitudes of chromospheric activity and magnetic field variability show clear dependence on the mean chromospheric activity and magnetic field strengths. However, the proportionality between $\log \Delta R'_{\rm HK}$ and mean $\log R'_{\rm HK}$ appears to undergo a change at around the middle of the main sequence, similar to the relation between the mean $\log R'_{\rm HK}$ and mean $\log |B_l|$. Stars with both high and low mean $\log R'_{\rm HK}$ can show similar levels of chromospheric activity variability.  

The distribution of $\log R^{\prime}_{\rm HK}$ data from our study, combined with data from \citet{Gomes2020} and \citet{BoroSaikia2018}, indicates a slightly under-populated region of intermediately--active, main sequence stars with spectral types from late-F to early-K and mean $\log R'_{\rm HK}$ between -4.5 and -4.75. The under-density of stars is not as distinct as the original Vaughan-Preston gap, which is consistent with the findings of \citet{BoroSaikia2018}. We do not find evidence for similar under-populated regions in the distributions of chromospheric activity variability, magnetic field strength nor magnetic field variability across our sample of stars. If the bimodal distribution of chromospheric activity we observed for late-F to early-K, main sequence stars is real, the fact that we do not see similar distributions in the magnetic field strength or chromospheric activity variability across our sample supports a change in dynamo properties on the main sequence, rather than a period of rapid stellar evolution. 

The ZDI data reveal that young stars are able to produce both dominantly poloidal fields and generate strong toroidal fields, like the most active main sequence stars. As stars spin down on the main sequence, mean chromospheric activity and activity variability amplitudes decrease as the large-scale magnetic field becomes more dominantly poloidal. Our results are consistent with mid-to-late F stars, G and K stars losing their ability to generate dominantly toroidal fields at estimated rotation periods of $\sim7$, 12 and 16 d respectively. For G-type stars, this occurs at $\log R'_{\rm{HK}}\sim-4.5$, and may roughly coincide with changes in the relationships between mean chromospheric activity, activity variability amplitudes and mean large-scale magnetic field strengths.

\section*{Acknowledgements}

We thank the referee for their constructive and thorough review, which has helped us to improve the quality of this manuscript.

This work is based on observations obtained at the TBL and CFHT. The TBL is operated by the Institut National des Sciences de l'Univers of the Centre National de la Recherche Scientifique of France (INSU/CNRS). The CFHT is operated by the the National Research Council of Canada, the INSU/CNRS and the University of Hawaii. We thank the staff at the TBL and CFHT for their time and data. The observations at the CFHT were performed with care and respect from the summit of Maunakea which is a significant cultural and historic site. We also acknowledge the use of the PolarBase database, which makes TBL and CFHT observations publicly available, and is operated by the CNRS, Observatoire Midi-Pyrénées and Université Toulouse III - Paul Sabatier. 

We acknowledge use of the {\sc{simbad}} and {\sc{VizieR}} data bases operated at CDS, Strasbourg, France. This work has also made use of the VALD database, operated at Uppsala University, the Institute of Astronomy RAS in Moscow, and the University of Vienna.

ELB is supported by an Australian Postgraduate Award Scholarship. SVJ acknowledges the support of the German Science Foundation (DFG) priority program SPP 1992 `Exploring the Diversity of Extrasolar Planets' (JE 701/5-1). AAV acknowledges funding from the European Research Council (ERC) under the European Union's Horizon 2020 research and innovation programme (grant agreement No 817540, ASTROFLOW). MMJ acknowledges funding from STFC consolidated grant ST/M001296/1.  SBS acknowledges the support of the Austrian Science Fund (FWF) Lise Meitner project M2829-N. VS acknowledges funding from the European Research Council (ERC) under the European Unions Horizon 2020 research and innovation programme (grant agreement No. 682393 AWESoMeStars) and support from the European Space Agency (ESA) as an ESA Research Fellow.

\section*{Data availability}

All NARVAL and ESPaDOnS data presented here are publicly available through the PolarBase data base (http://polarbase.irap.omp.eu/). Archival chromospheric activity data used for this study are available via {\sc{vizier}}.


\DeclareRobustCommand{\DO}[3]{#3}
\DeclareRobustCommand{\DE}[3]{#3}

\bibliographystyle{mnras}
\bibliography{library.bib} 

\appendix

\section{Significance of $\log R'_{\rm HK}$ activity distributions}

\begin{figure}
    \centering
    \includegraphics[width=\columnwidth]{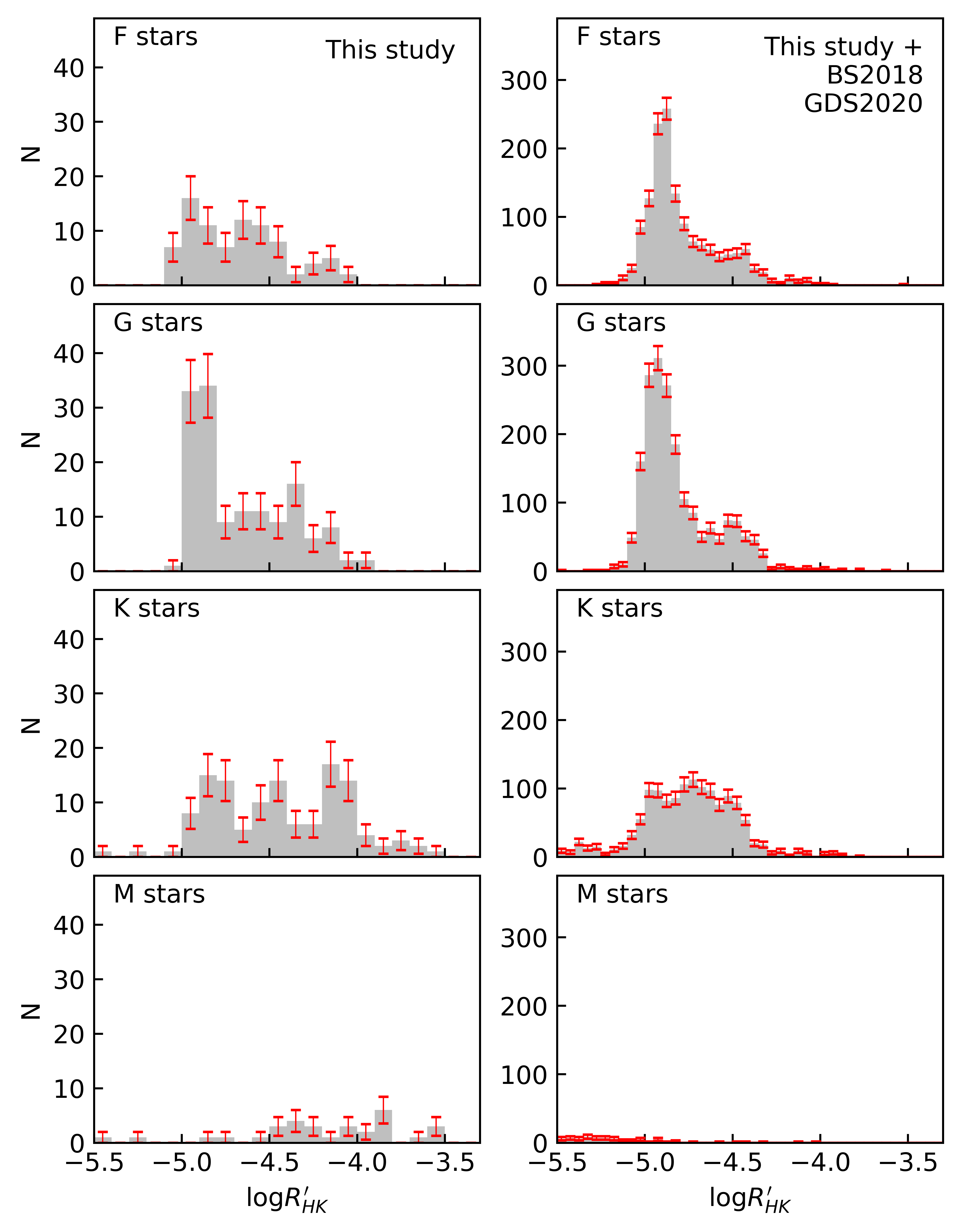}
    \caption{Histograms showing the distributions of $\log R'_{HK}$ activity across F, G, K and M stars, with error bars showing the $1\sigma$ uncertainty for each bin. }
    \label{fig:AppA}
\end{figure}

Figure \ref{fig:AppA} shows the distributions of $\log R'_{\rm HK}$ activity across samples of F, G, K and M stars, including both young and mature stars. On the left we show the data from our study only, and on the right we combine our data with the publicly accessible $\log R'_{\rm HK}$ measurements of \citet{Gomes2020}, and activity data compiled by \citet{BoroSaikia2018} from the Mount Wilson Project \citep{Duncan1991,Baliunas1995}, HARPS \citep{Lovis2011,Bonfils2013}, Lowell \citep{Hall2009}, CPS \citep{Wright2004,Isaacson2010}, Magellan \citep{Arriagada2011} and Southern Stars \citep{Henry1996,Gray2005} surveys. The combined sample is made up of 5179 stars. Both the \citet{BoroSaikia2018} and \citet{Gomes2020} data include stellar B-V, which we used to convert their mean S-indices to $R^{\prime}_{HK}$ values using the method described in section \ref{sec:S-index}.  \citet{Gomes2020} provides stellar effective temperatures, and for the \citet{BoroSaikia2018} data we converted B-V to $T_\mathrm{eff}$ according to \citet{Ballesteros2012}. We note that since the stellar B-V and temperature were taken from the published data tables, and were not calculated consistently with section \ref{sec:params_from_MIST}, they will have some impact on the data distributions for the combined sample.  The error bars in Figure \ref{fig:AppA} indicate the $1\sigma$ uncertainties in the height of each histogram bin, which we determined from the Poisson uncertainty ($\sqrt{N}$). They indicate that a bimodal distribution is significant with respect to the $1\sigma$ uncertainties for the combined sample of G stars. For F and K stars in the combined sample, bimodal/triple peaked activity distributions are not statistically significant.  

\section{Mean chromospheric activity, magnetic field strength and estimated rotation period for G stars}

\begin{figure}
    \centering
    \includegraphics[width=0.75\columnwidth]{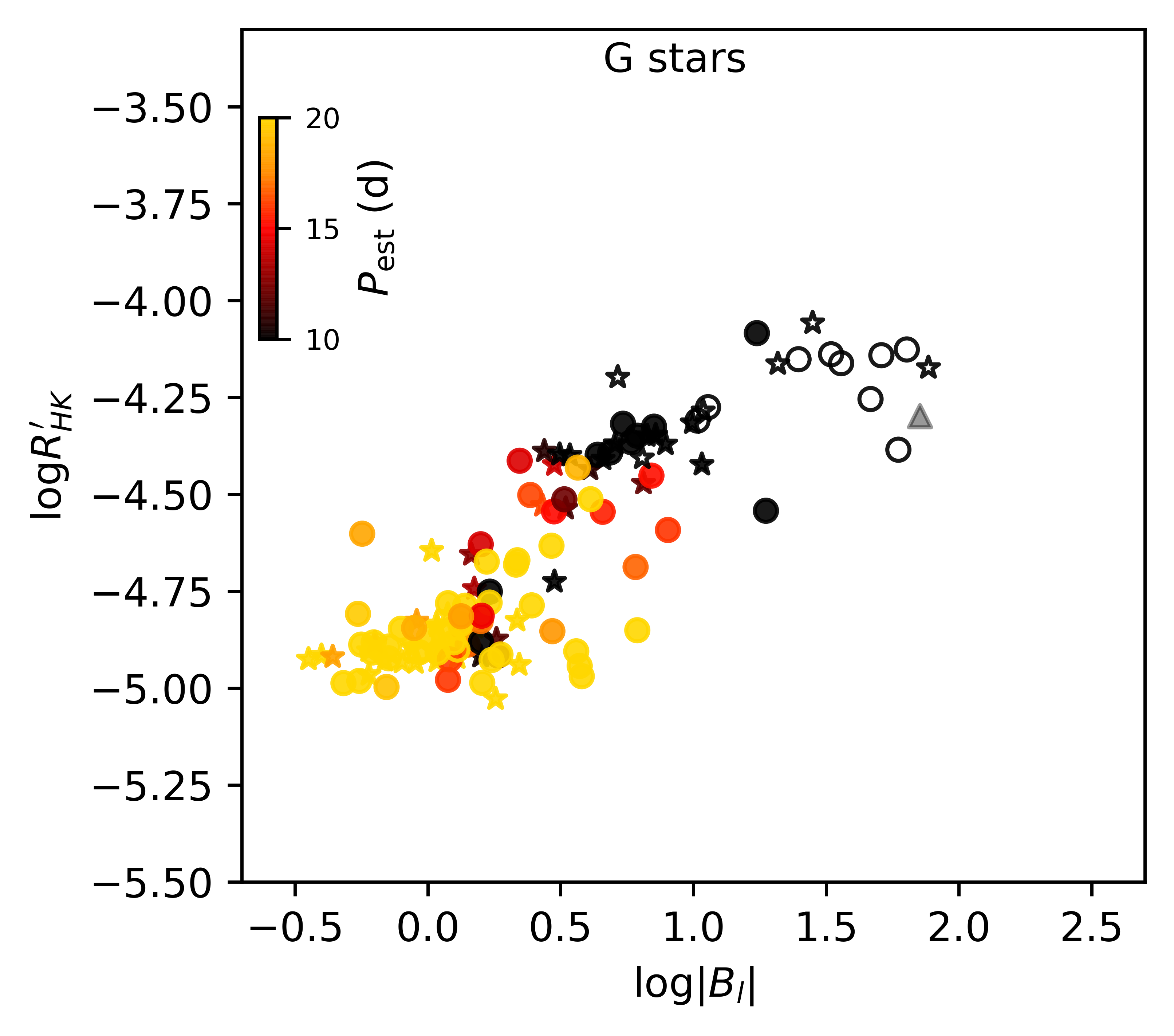}
    \caption{Mean chromospheric activity versus mean longitudinal magnetic field strength for G stars, where color scales with the estimated rotation period, $P_{\mathrm{est,} i=60\degr}$. Marker shapes/styles are the same as in Figure \ref{fig:HRD}.}
    \label{fig:AppB}
\end{figure}

Figure \ref{fig:AppB} shows the same activity and magnetic field data as Figure \ref{fig:SvBl} for G stars, but marker colour scales with the estimated rotation period, $P_{\mathrm{est,} i=60\degr}$. The `step down' in chromospheric activity at around $\log|B_l|\sim0.4$ occurs as G stars reach estimated rotation periods of around 15d. 

\section{Chromospheric activity variability data from Gomez da Silva et al. (2020)}

\begin{figure}
    \centering
    \includegraphics[width=0.7\columnwidth]{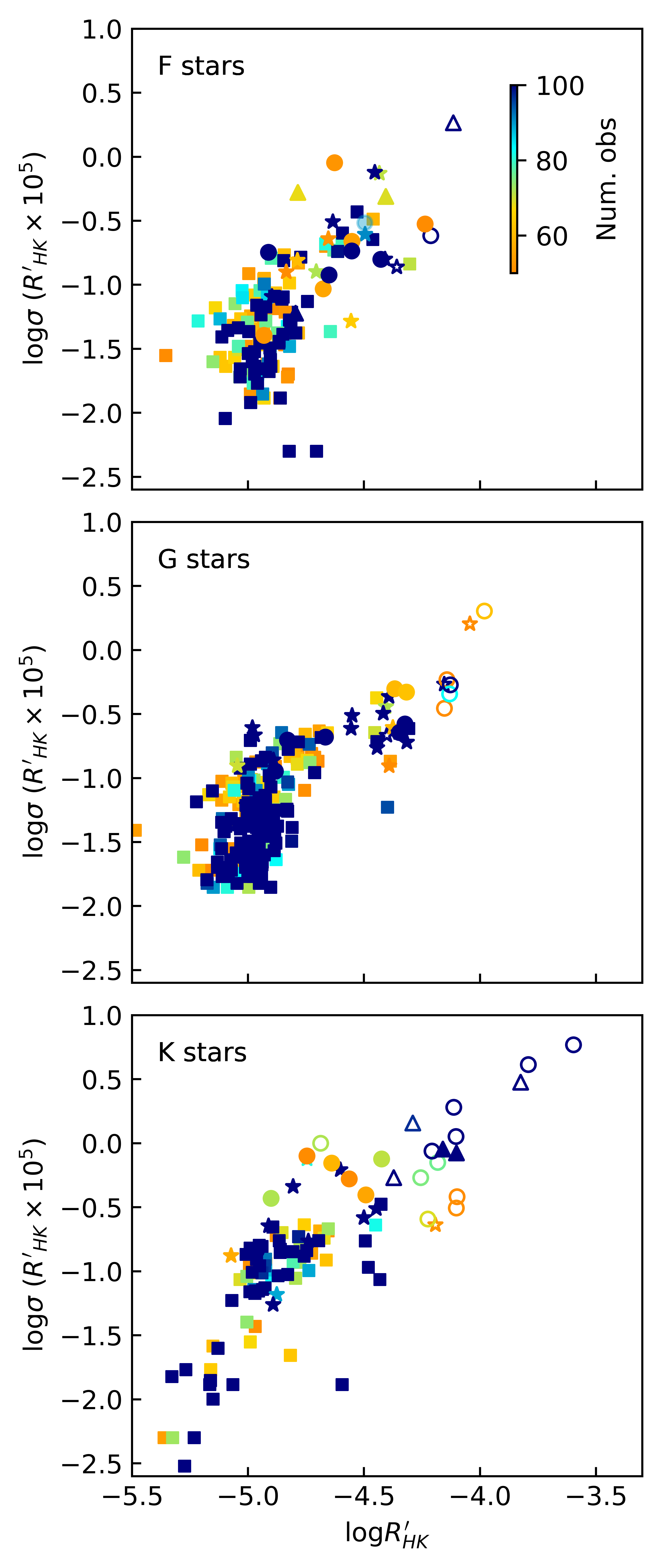}
    \caption{Activity variability versus mean activity level for our data and the HARPS sample studied by \citet{Gomes2020}. Squares indicate data from \citet{Gomes2020}, while circles, stars and triangles show our data, with marker styles the same as in Figure \ref{fig:HRD}. For stars common to both studies we have shown our results only. We have filtered the data to show only stars with $\geq50$ observations, and marker color scales with the number of observations. }
    \label{fig:AppC}
\end{figure}

\citet{Gomes2020} studied chromospheric activity and activity variability amplitudes for HARPS planet search stars, and we combine their data with our results in Figure \ref{fig:AppC}. The method used by \citet{Gomes2020} to convert S-indices to $R'_{HK}$ values was different compared to our study.  While we use the photospheric and bolometric correction factors from \citet{Suarez_Mascareno2015} due to the large B-V range of our sample, \citet{Gomes2020} derived photospheric contributions according to \citet{Noyes1984} and bolometric corrections from \citet{Rutten1984}. For Figure \ref{fig:AppC} we have adapted our $R'_{HK}$ conversion to follow \citet{Gomes2020}. Rather than showing the peak-to-peak activity variability amplitudes we present in section \ref{sec:var_amplitudes}, Figure \ref{fig:AppC} shows the standard deviation of $R'_{HK} \times 10^5$ for each target, consistent with \citet{Gomes2020}. 

\section{Chromospheric activity and magnetic field geometry versus Rossby number}

\begin{figure}
    \centering
    \includegraphics[width=0.9\columnwidth]{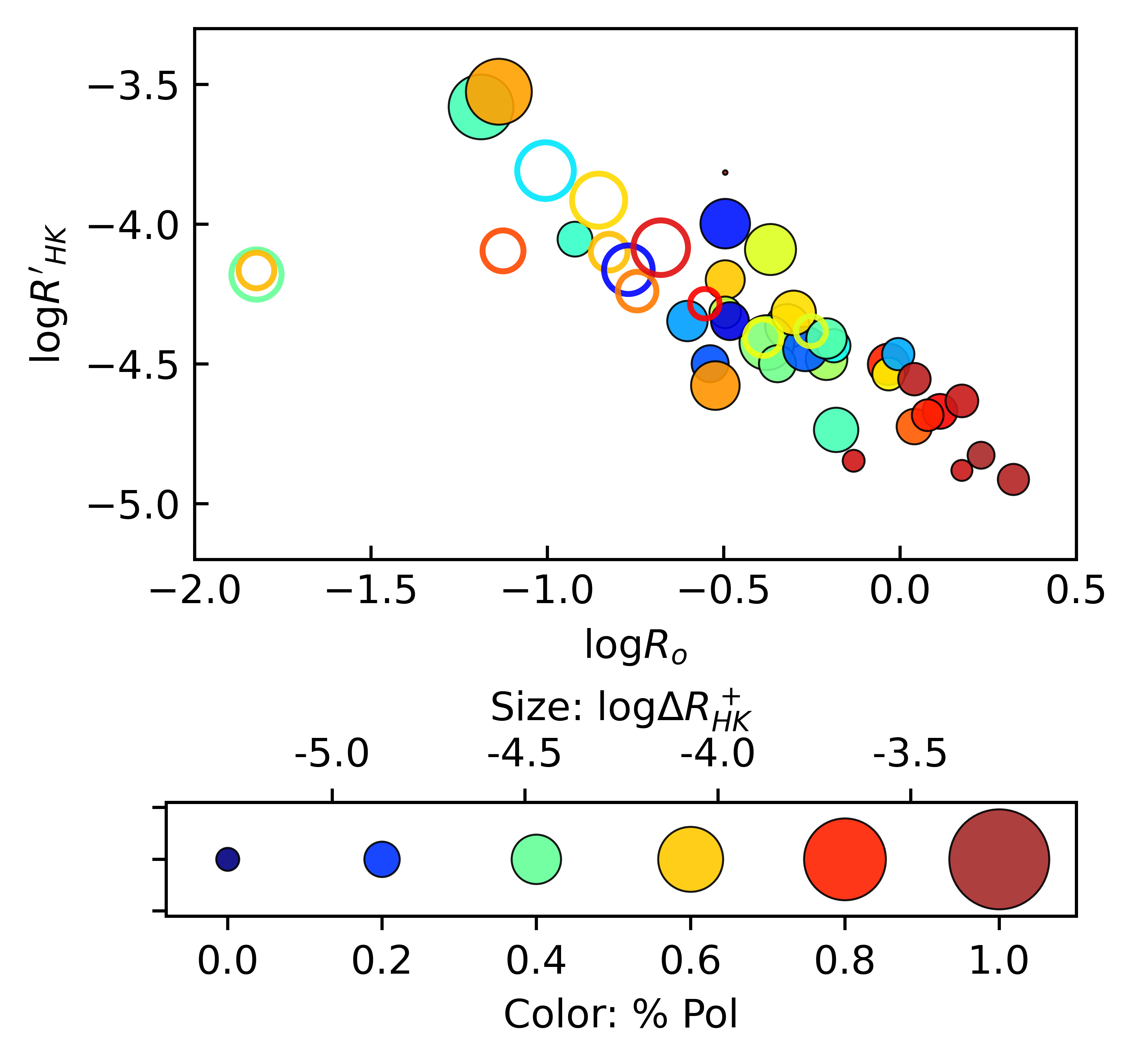}
    \caption{Mean $\log R^{\prime}_{\rm HK}$ versus the theoretical Rossby number taken from \citet{See2019}, where marker colour shows the fractional poloidal field and marker size scales with chromospheric activity variability.}
    \label{fig:AppD}
\end{figure}

Figure \ref{fig:AppD} compares chromospheric activity, activity variability (symbol size) and magnetic field geometry (symbol colour) to the theoretical Rossby number, taken from \citet{See2019}. F, G, K and M stars are shown together, with main sequence stars indicated by filled circles and young stars indicated by open circles. Chromospheric activity variability is reduced and the fraction of the magnetic field stored in the poloidal component becomes stronger as $\log Ro$ increases. All stars with $\log Ro \geq 1$ show dominantly poloidal magnetic field geometries.

\bsp	
\label{lastpage}
\end{document}